\documentclass[10pt,conference]{IEEEtran}
\usepackage{cite}
\usepackage{amsmath,amssymb,amsfonts}
\usepackage{algorithmic}
\usepackage{graphicx}
\usepackage{textcomp}
\usepackage{xcolor}
\usepackage[hyphens]{url}
\usepackage{fancyhdr}
\usepackage{hyperref}

\usepackage{amsmath,amssymb,amsfonts}
\usepackage{comment}
\usepackage{gensymb}
\usepackage{textcomp}
\usepackage{xcolor}
\usepackage[hyphens]{url}
\usepackage{booktabs} 
\usepackage[utf8]{inputenc} 
\usepackage[T1]{fontenc}    
\usepackage{url}            
\usepackage{booktabs}       
\usepackage{nicefrac}       
\usepackage[export]{adjustbox}
\usepackage{comment}
\usepackage{booktabs} 
\usepackage{algorithmic}
\usepackage[vlined,ruled]{algorithm2e}
\usepackage{amsfonts}       
\usepackage{nicefrac}       
\usepackage{microtype}      
\usepackage{verbatim}
\usepackage{caption}
\usepackage{bbm}
\usepackage{times}
\usepackage{epsfig}
\usepackage{graphicx}
\usepackage{amsmath}
\usepackage{amssymb}
\usepackage{amsthm}
\usepackage{multirow}
\usepackage{bbm}
\usepackage{hyperref}
\usepackage{multicol}
\usepackage{algorithmic}
\usepackage{caption}%
\usepackage[affil-it]{authblk}

\newcommand{\squeezeup}{\vspace{-3mm}}
\usepackage[normalem]{ulem}

\newcommand{\hpcayear}{2024}

\title{CAMEL: Co-Designing AI Models and eDRAMs for Efficient On-Device Learning}

\author[1]{Sai Qian Zhang\IEEEauthorrefmark{1}\IEEEauthorrefmark{2}}
\thanks{The first author wants to thank someone.}

\author[2]{Thierry Tambe\IEEEauthorrefmark{1}}
\author[2]{Nestor Cuevas}
\author[2]{Gu-Yeon Wei}
\author[2]{David Brooks}
\affil[1]{Department of Electrical Engineering and Department of Computer Science, New York University}
\affil[2]{John A. Paulson School of Engineering and Applied Sciences, Harvard University}

\fancypagestyle{camerareadyfirstpage}{%
  \fancyhead{}
  
  \fancyhead[C]{
    \ifdefined\aeopen
    \parbox[][12mm][t]{13.5cm}{\hpcayear{} IEEE International Symposium on High-Performance Computer Architecture (HPCA)}    
    \else
      \ifdefined\aereviewed
      \parbox[][12mm][t]{13.5cm}{\hpcayear{} IEEE International Symposium on High-Performance Computer Architecture (HPCA)}
      \else
      \ifdefined\aereproduced
      \parbox[][12mm][t]{13.5cm}{\hpcayear{} IEEE International Symposium on High-Performance Computer Architecture (HPCA)}
      \else
      \parbox[][0mm][t]{13.5cm}{\hpcayear{} IEEE International Symposium on High-Performance Computer Architecture (HPCA)}
    \fi 
    \fi 
    \fi 
    \ifdefined\aeopen 
      \includegraphics[width=12mm,height=12mm]{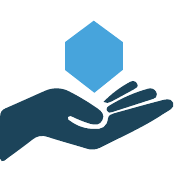}
    \fi 
    \ifdefined\aereviewed
      \includegraphics[width=12mm,height=12mm]{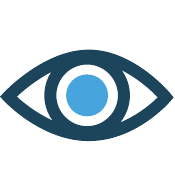}
    \fi 
    \ifdefined\aereproduced
      \includegraphics[width=12mm,height=12mm]{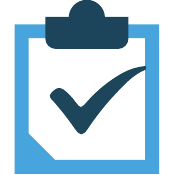}
    \fi
  }
  \fancyfoot[C]{}
}
\begin{document}
\maketitle

\ifdefined\hpcacameraready 
  \thispagestyle{camerareadyfirstpage}
  \pagestyle{empty}
\else
  \thispagestyle{plain}
  \pagestyle{plain}
\fi

\newcommand{\hpcaheight}{0mm}
\ifdefined\eaopen
\renewcommand{\hpcaheight}{12mm}
\fi


\begin{abstract}
On-device learning\let\thefootnote\relax\footnotetext{\IEEEauthorrefmark{1} Equal contribution.} \let\thefootnote\relax\footnotetext{\IEEEauthorrefmark{2} The work was conducted while the author was affiliated with Harvard University.}allows AI models to adapt to user data, thereby enhancing service quality on edge platforms. However, training AI on resource-limited devices poses significant challenges due to the demanding computing workload and the substantial memory consumption and data access required by deep neural networks (DNNs). To address these issues, we propose utilizing embedded dynamic random-access memory (eDRAM) as the primary storage medium for transient training data. In comparison to static random-access memory (SRAM), eDRAM provides higher storage density and lower leakage power, resulting in reduced access cost and power leakage. Nevertheless, to maintain the integrity of the stored data, periodic power-hungry refresh operations could potentially degrade system performance.

To minimize the occurrence of expensive eDRAM refresh operations, it is beneficial to shorten the lifetime of stored data during the training process. To achieve this, we adopt the principles of algorithm and hardware co-design, introducing a family of reversible DNN architectures that effectively decrease data lifetime and storage costs throughout training. Additionally, we present a highly efficient on-device training engine named \textit{CAMEL}, which leverages eDRAM as the primary on-chip memory. This engine enables efficient on-device training with significantly reduced memory usage and off-chip DRAM traffic while maintaining superior training accuracy. We evaluate our CAMEL system on multiple DNNs with different datasets, demonstrating a $2.5\times$ speedup of the training process and $2.8\times$ training energy savings than the other baseline hardware platforms.

\end{abstract}

\section{Introduction}

The looming deluge of data is expected to shift AI-related workloads to the edge and end devices~\cite{src_decadal_plan}. Processing data at the edge comes with numerous benefits, such as better latency, enhanced energy efficiency, increased security and privacy, and improved autonomy. In particular, there is a growing demand for training DNNs locally within the edge device. For example, Federated Learning (FL) requires on-device DNN training with local user data, which enables users to adapt the DNN model with personal data and continuously improve their accuracy based on users' preference. 

\begin{figure}
    \centering\begin{adjustbox}{width=1\columnwidth,center}
    \includegraphics[width=0.45\textwidth]{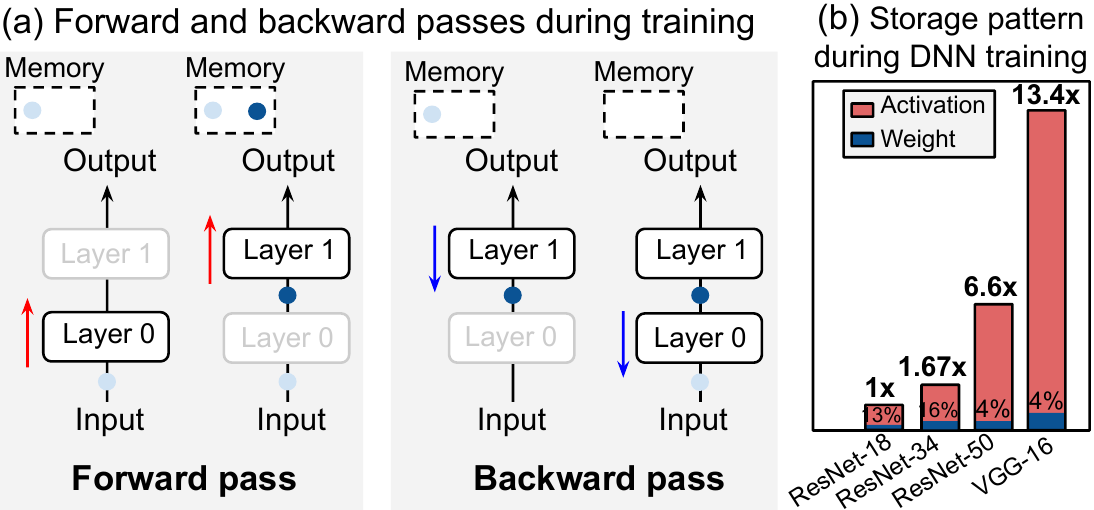}
    \end{adjustbox}
    \caption{(a) A single iteration of two-layer DNN training. (b) Normalized size of activation and weight (shown in percentage of the total size) during a training iteration. The batch size is set to 48.}
    \label{fig:dnn-training-ex}
    \squeezeup
\end{figure}

DNN training involves tuning the network's weights and biases to achieve high validation accuracy. It starts with random learnable parameters that are iteratively updated through training iterations. Each iteration processes a small dataset subset (mini-batch) in two steps: forward pass to compute the training loss and backward pass to calculate gradients for updating the DNN weights. Figure~\ref{fig:dnn-training-ex}(a) illustrates the forward pass for a two-layer DNN, where input activations are multiplied with layer weights to generate output activations. The input activations of all layers need to be buffered in memory for later gradient computations.
In the backward pass, the output gradient and the previous layer's input activation are multiplied to obtain the weight gradient of the current layer. This gradient is then added to the current weight for the update. The output gradient is multiplied with the weight to get the preceding layer's output gradient. This process repeats until all weights are updated.

Training DNNs on edge compute platforms presents numerous challenges. Compared to inference, training imposes much higher compute requirements due to additional calculations for activation and weight gradients. Memory requirements are also much higher. All intermediate activations computed during the forward pass must be buffered in memory to later compute the gradients. Figure~\ref{fig:dnn-training-ex}(b) illustrates the sizes of activations and weights that need to be stored during training for ResNet-18, ResNet-34, ResNet-50, and VGG-16 on the CIFAR-10 dataset. Activations consume most of the memory space, approximately 13 times larger than the weights on average. On top of this, small edge devices typically have limited on-chip storage, leading to frequent and costly accesses to off-chip memories. These high compute and memory requirements have traditionally discouraged large-scale DNN training on resource-constrained edge devices.

Yet, the unmitigated growth of deep learning models used in all forms of applications motivates us to tackle the aforementioned challenges to local on-device training. First, we propose using embedded DRAM (eDRAM) as the main on-chip medium to store activations and gradients computed during DNN training. Compared with SRAM, eDRAM has the innate potential for higher data storage density by, for example, utilizing fewer transistors in a memory cell, e.g., 3T DRAM cell versus 6T SRAM cell ($>2\times$~\cite{giterman20201, chen2014dadiannao}). Also, the 3T cell topology eliminates the VDD-to-GND leakage paths inherent to 6T SRAM cells, offering much lower leakage power ($3.5\times$ lower according to~\cite{chun2011edram}). While conventional 1T1C DRAM cells are even denser than 3T cells, they require an entirely different process technology required to implement the high density capacitors. Furthermore, 3T cells can be as fast as 6T SRAM cells given they retain one side of the SRAM cell's read path transistors. Hence, the benefits of eDRAM make it an attractive option for DNN computing tasks~\cite{chen2014dadiannao,tu2018rana,nguyen2019st}. 

However, the main drawback of using eDRAMs is the inherent need to periodically refresh the cells, lest they lose data due to leakage. Additionally, unlike 1T1C cells that can be refreshed just by reading the cells. 3T cells require a read followed by a write to explicitly refresh the cell, imposing high latency and power penalties that hamper on-chip training~\cite{bhati2015dram, giterman20201}. This is where co-design comes to the rescue. We propose an algorithm-hardware co-design strategy that modifies the DNN model architecture in light of the constraints imposed by using eDRAM in order to reap the associated benefits, i.e., higher on-chip memory density.
%

\begin{figure}
    \centering
    \includegraphics[width=0.5\textwidth]{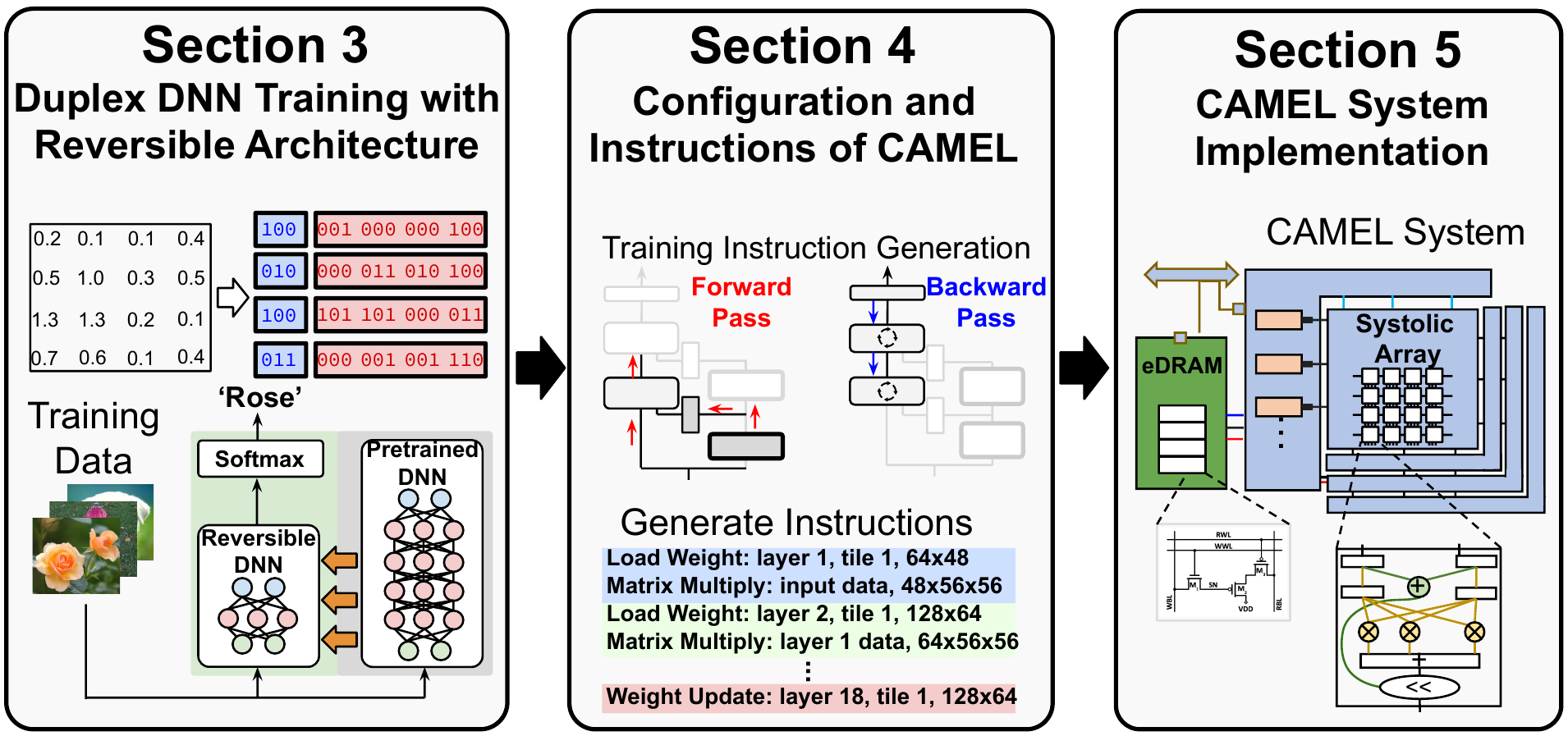}
    \caption{An overview of the CAMEL.}
    \label{fig:overview}
    \squeezeup
\end{figure}

We propose a family of reversible DNN architectures that first reduces active storage requirements overall and also reduces data lifetime requirements potentially below the refresh interval of the eDRAM, further eliminating the refresh penalty. Activation data is largely transient, generated and consumed within a single training iteration. 
Recall, training traditionally requires holding on to all activations computed during the forward pass. Moreover, during the backward pass operation, activations in the early layers must be stored longer than those in later layers, as depicted by the light blue and dark blue dots in Figure~\ref{fig:dnn-training-ex}(a). Reversible DNN architectures do not require holding on to the intermediate activations computed during the forward pass by instead recomputing values as needed during the backward pass. In essence, reversibility alleviates memory requirements (both footprint and lifetime) at the cost of more compute. Given current CMOS technology scaling trends, this is a reasonable trade. 

To support this co-design strategy that leverages the mutual and synergistic benefits offered by combining eDRAMs with reversible DNN architectures, we introduce an efficient DNN training engine we call CAMEL. eDRAMs offer higher on-chip memory density but suffer refresh penalties. Reversible architectures minimize transient data storage in terms of size and duration, eliminating need for eDRAM refresh and expensive off-chip DRAM traffic, all the while maintaining superior training accuracy. Our contributions are summarized as follows (Figure~\ref{fig:overview}):
 

\begin{itemize}
\item We introduce a novel family of DNN architectures designed for efficient DNN training, eliminating the need to store intermediate activations. This solution, named \textit{Duplex DNN (DuDNN)}, involves a straightforward modification to the DNN forward rule, resulting in a substantial reduction in data lifetime and significantly reducing the frequency of costly eDRAM refresh operations (Section~\ref{sec:camel-DuDNN}).
\item We describe an efficient computation pattern for both forward and backward passes of DuDNN training. The proposed computation pattern can achieve optimal memory footprint and data lifetime, leading to significant savings in terms of memory traffic and space (Section~\ref{sec:computation-pattern}). 
\item We design an efficient DNN training hardware engine with eDRAM as the main on-chip storage medium. To promote greater energy efficiency, the specialized accelerator supports a novel processing element (PE) with two levels of computation gating (Section~\ref{sec:camel-hardware}).
\item We extensively evaluate CAMEL using diverse DNN architectures, training datasets, and system configurations. Evaluation results confirm CAMEL achieves an impressive reduction of more than $2.5 \times$ in training latency and a significant training energy reduction of $2.8\times$.
\end{itemize}

\section{Background and Related Work}
In Section~\ref{sec:bg:dnn-training}, we introduce the matrix operations within the DNN training process. Section~\ref{sec:reversible-dnn-intro} provides an overview of reversible DNN architecture and its application on DNN training. In Section~\ref{sec:bg:bfp}, we describe the different numerical formats used for DNN training, followed by a literature review on DNN training accelerator in Section~\ref{sec:dnn-accelerator-review}. In Section~\ref{sec:edram-intro}, we review the memory architectures in DNN accelerators.



\subsection{Computations in DNN Training}
\label{sec:bg:dnn-training}
\begin{table*}
    \centering
    \caption{The forward and backward pass steps for a single layer of DNN training represented in a convolution view and a matrix view. The kernel size of the weight filters is assumed to be $1\times 1$ for illustration simplicity.}
    \includegraphics[width=\textwidth]{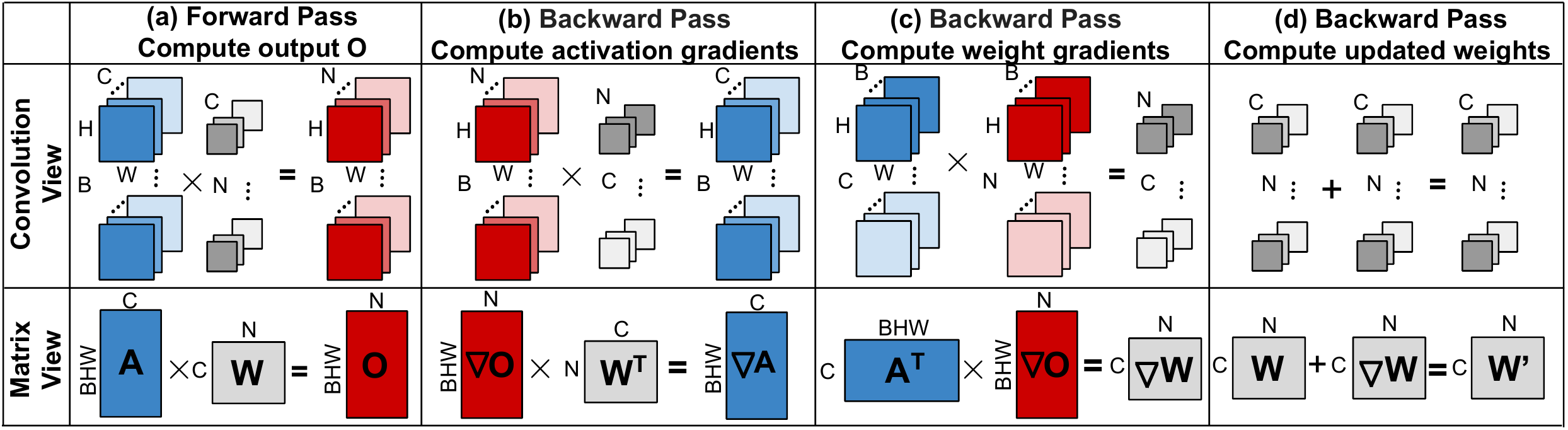}
    \label{tab:dnn-training-computation}
    \squeezeup
\end{table*}
DNN training consists of a series of iterations, each involving a forward pass and a backward pass. In the forward pass, a mini-batch of data is input into the DNN to compute the training loss. The input activations ($A$) are convolved with the layer weights ($W$) to generate the output ($O$), which is then processed through the batch normalization and activation functions. The resulting intermediate values are used as inputs for subsequent layers. In the backward pass, gradients are calculated from the training loss and used to update the DNN weights. Specifically, the output gradient ($\nabla O$) is convolved with the weights ($W$) to obtain the input gradient ($\nabla A$), and then convolved with the input activations ($A$) to generate the weight gradient ($\nabla W$). The weight gradient is multiplied by the learning rate ($\eta$) and added to the original weights ($W$) in an elementwise manner to obtain the updated weights ($W'$). Additional operations are required for other types of optimizers such as Adam~\cite{kingma2014adam}. Table~\ref{tab:dnn-training-computation} provides details on the tensor operations and equivalent matrix computations involved in the forward and backward passes for a convolutional (CONV) layer.

\subsection{Reversible DNN Architecture}
\label{sec:reversible-dnn-intro}
\begin{figure}
    \centering
    \includegraphics[width=0.48\textwidth]{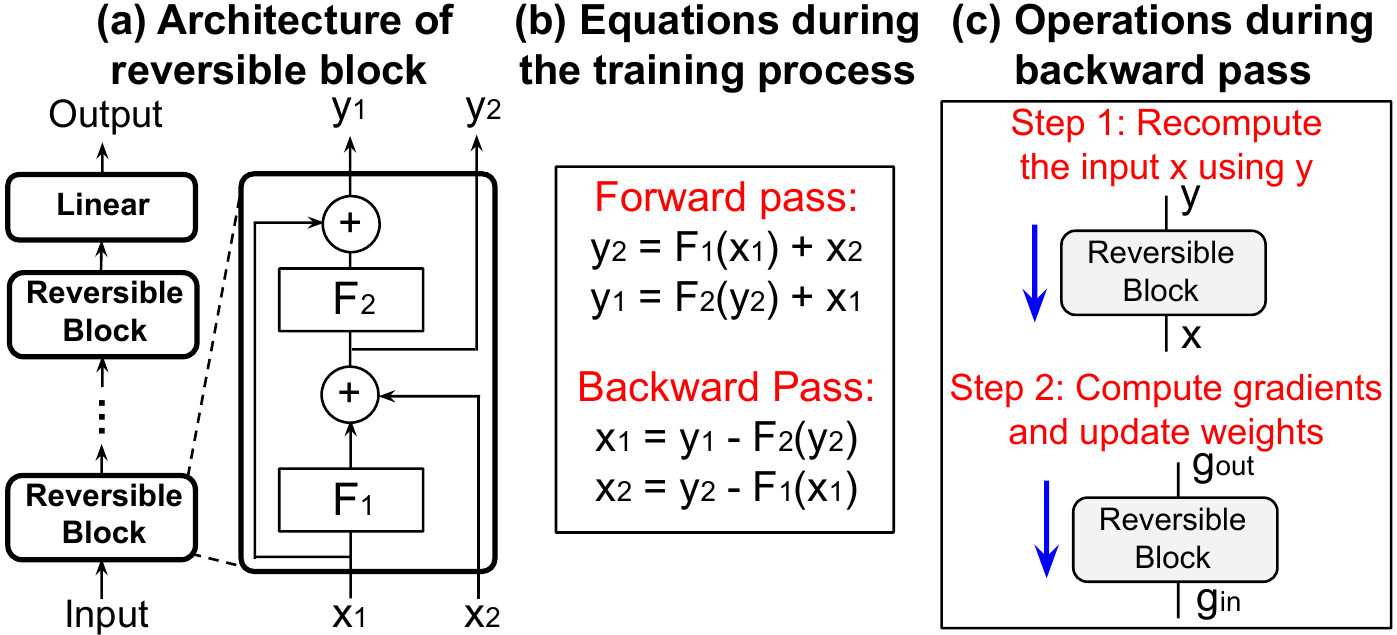}
    \caption{Reversible DNN architecture and computations.}
    \label{fig:rev-arch-ex}
    \squeezeup
\end{figure}

A reversible residual network (RevNet)~\cite{gomez2017reversible} is a variant of the canonical residual neural network (ResNet)~\cite{he2016deep}. RevNet comprises multiple blocks that are reversible in the sense that the input activation of the block can be recovered using its output activation. The architecture of a reversible block is illustrated in Figure~\ref{fig:rev-arch-ex}(a), where $F_{1}$ and $F_{2}$ represent a series of DNN layers. In the case of CNN-based architectures, $F_{1}$ and $F_{2}$ typically consist of a CONV layer, a batch normalization layer, and a ReLU layer. The reversible block takes two input activations, denoted as $x_{1}$ and $x_{2}$, and produces two output activations, $y_{1}$ and $y_{2}$, as shown in Figure~\ref{fig:rev-arch-ex}(b). Specifically, $y_{1}$ and $y_{2}$ can be computed as follows:
\begin{equation}
    \label{eqn:forward-eqn}
    y_{2} = x_{2} + F_{1}(x_{1}) \text{  and  } 
    y_{1} = x_{1} + F_{2}(y_{2})
\end{equation}
The following expressions can then recompute $x_{1}$ and $x_{2}$ from $y_{1}$ and $y_{2}$:
\begin{equation}
    \label{eqn:backward-eqn}
    x_{1} = y_{1} - F_{2}(y_{2}) \text{  and  } 
    x_{2} = y_{2} - F_{1}(x_{1}) 
\end{equation}
The reversible architecture enables the backward pass computations to be performed without the need to store the input activations, as all the input activations $x$ can be recomputed using the output $y$. Figure~\ref{fig:rev-arch-ex}(c) depicts the operations involved during the backward pass of a reversible DNN layer. Given the output $y$, the input activations are first recomputed with Equation~\ref{eqn:backward-eqn}. Afterwards, the input and weight gradients are computed with standard backward pass operations. Hence, this reversibility yields significant memory savings and reduction in data lifetime. 

Since the first introduction of the residual architecture in~\cite{he2016deep}, residual DNNs have been widely adopted across a wide range of applications, including image recognition~\cite{he2016deep, dosovitskiy2020image, he2022masked}, object detection~\cite{redmon2018yolov3,wang2022internimage, he2022masked}, image segmentation~\cite{hatamizadeh2022unetr, kirillov2023segment,zhang2023faster}, natural language processing~\cite{devlin2018bert, vaswani2017attention, brown2020language, touvron2023llama}, etc. 
The residual architecture has gained widespread use given its ability to train deep networks with many layers, enabling significant advancements in accuracy for various demanding tasks. 
Consequently, we expect residual architectures to maintain their relevance in future DNN developments. 

\begin{figure}
    \centering
    \includegraphics[width=0.48\textwidth]{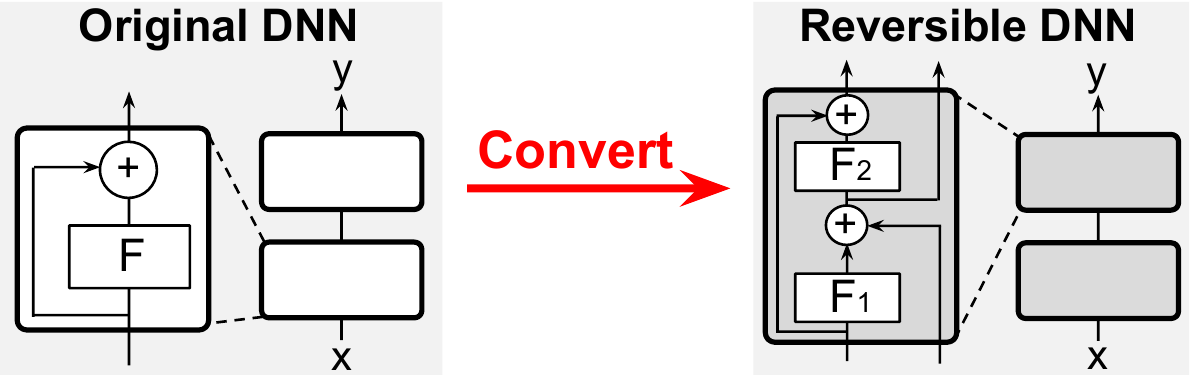}
    \caption{Converting a standard DNN into DuDNN.}
    \label{fig:conversion}
    \squeezeup
\end{figure}

Residual models (such as ResNet and transformer) can be made reversible with minimal modifications to their forwarding rule. 
As an illustrative example, let's use a two-layer DNN shown in Figure~\ref{fig:conversion}. The residual block, denoted as F, corresponds to a stack of layers (e.g., CONV+BN+ReLU). By inserting a scaled version of F (i.e., $F_{1}$, $F_{2}$) into the architecture, the DNN can be made reversible. In the evaluation section, we show that the reversible DNN can achieve similar accuracies for various types of F.

\subsection{Number Formats for Efficient DNNs Training}
\label{sec:bg:bfp}

Various numerical formats have been explored to reduce the computing load in DNN training. Fixed-point formats~\cite{courbariaux2015binaryconnect,zhu2016trained,hubara2017quantized,park2017weighted,kapur2017low,banner2018scalable,jacob2018quantization,mcdanel2019full, kung2020term,han2021hnpu, zhang2021training} and floating-point formats~\cite{micikevicius2018mixed,tf32,bfloat,sun2019hybrid,zhang2022fast,ttambe2020dac} have been extensively studied for efficient DNN training and inference in research literature. In contrast, the Block Floating-Point (BFP)~\cite{Wilkinson_64} format offers a middle ground between floating-point and fixed-point formats. BFP groups together several values with a common group exponent while retaining individual mantissas. 

\begin{figure}
    \centering
    \includegraphics[width=0.48\textwidth]{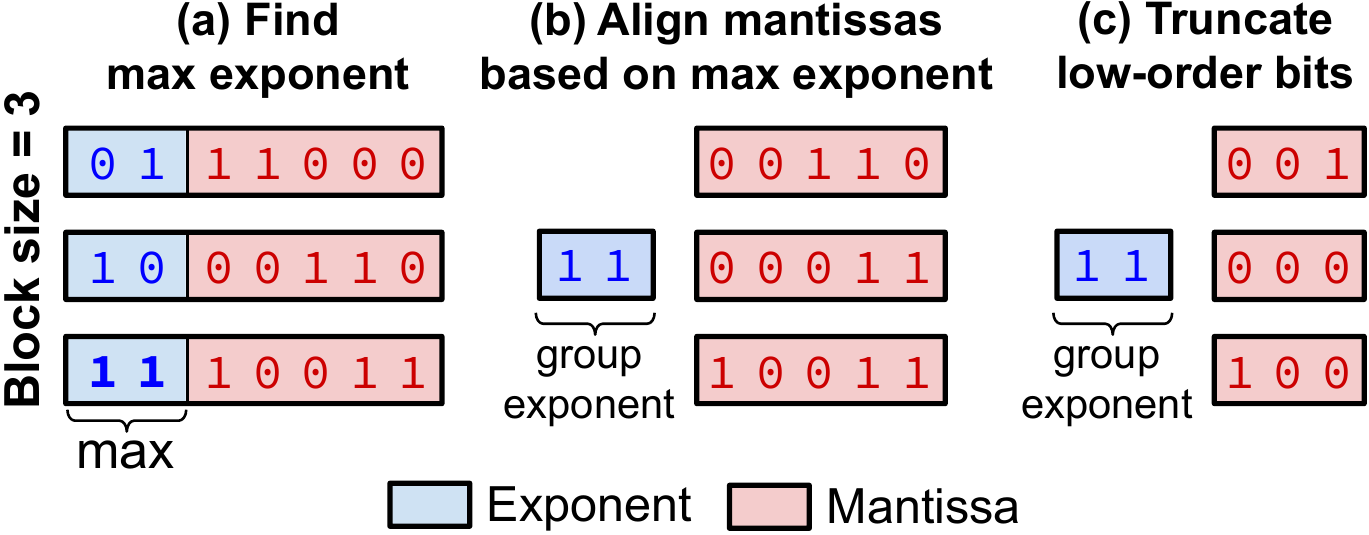}
    \caption{An example on BFP quantization.}
    \label{fig:bfp-format}
    \squeezeup
\end{figure}

Figure~\ref{fig:bfp-format} illustrates the BFP conversion process for a group of three floating-point numbers. The largest exponent in the group becomes the shared exponent, and the mantissas of each number are aligned based on the difference between the group exponent and their individual exponents (Figure~\ref{fig:bfp-format}(b)). The aligned mantissas are then truncated to a lower bitwidth (Figure~\ref{fig:bfp-format}(c)) for efficient operation. BFP achieves higher computing efficiency than floating-point formats while maintaining a higher dynamic range and lower quantization error compared to fixed-point formats. The multiplication operations between two BFP groups can be performed using fixed-point arithmetic without aligning the mantissas, as all numbers in a BFP group share the same exponent. Lastly, exponent addition only needs to be performed once between two group exponents.

\subsection{Accelerator Designs for DNN Training }
\label{sec:dnn-accelerator-review}
Prior research has explored diverse techniques to accelerate DNN training using sparsity in weights and activations~\cite{mahmoud2020tensordash, zhang2019eager, yang2020procrustes, choi2020energy, qin2020sigma, zhang2022fast}. For example, Procrustes~\cite{yang2020procrustes} and Eager Pruning~\cite{zhang2019eager} enhance training efficiency by co-designing algorithms to match hardware capabilities, eliminating unimportant DNN weights and improving hardware efficiency. TensorDash~\cite{mahmoud2020tensordash} accelerates DNN training by removing ineffective operations from sparse input activations. Another approach involves reducing DNN operand precision~\cite{judd2016stripes, lee20197}, such as using fine-grained mixed precision (FGMP)~\cite{lee20197} or dynamically adjusting precision during DNN training as proposed in FAST~\cite{zhang2022fast}. 

\subsection{Memory Architectures in DNN Accelerators}
\label{sec:edram-intro}

\begin{table}
    \centering
\includegraphics[width=\columnwidth]{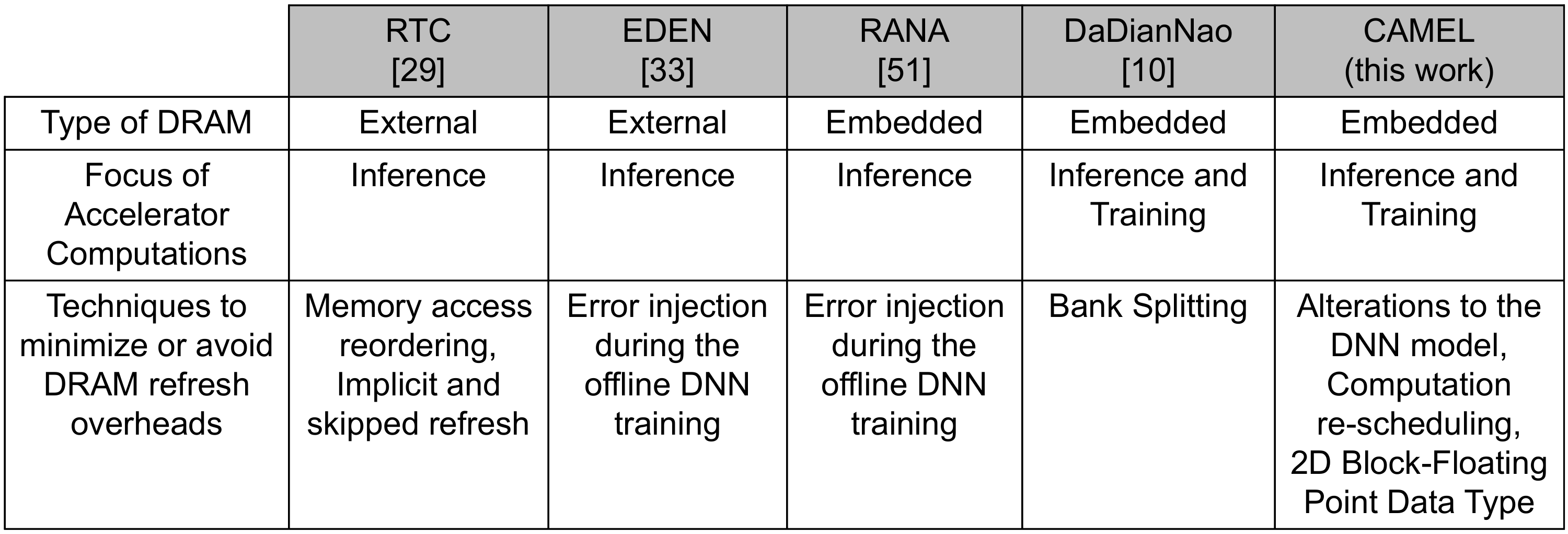}
    \caption{Qualitative comparison of CAMEL with prior related work based on DRAM and eDRAM.}
    \label{fig:prior_edram_work}

\end{table}
Memory access has been recognized as a major contributor to the overall power consumption in DNN accelerators~\cite{han2015deep,zheng2022optimizing}. To reduce the expensive off-chip DRAM accesses during DNN inference, previous literature has explored different types of on-chip SRAM such as 6T or 10T~\cite{zhang2016machine,biswas2018conv}.

Multiple eDRAM circuit topologies~\cite{yu2020logic,giterman20201}, using as few as 2 transistors, have been proposed as an alternative to SRAM.
Consequently, recent studies~\cite{tu2018rana,chen2014dadiannao} have explored the use of eDRAM in an accelerator system for efficient DNN computations. In addition, given the refresh overheads coming with the usage of dynamic memories, there has been efforts~\cite{rtc,eden,tu2018rana} towards mitigating or minimizing this undesirable cost. 
These prior work are qualitatively compared with CAMEL in TABLE~\ref{fig:prior_edram_work}. 
First, we note that most work~\cite{eden,rtc,tu2018rana} investigate DRAM compliance by co-designing the accelerator for inference computations. In contrast, our work further addresses the challenges of using eDRAM for on-chip DNN training. In addition, our work is leveraging eDRAM as the main storage medium of training parameters as opposed to external DRAM (as done in EDEN~\cite{eden} and RTC~\cite{rtc}).
Furthermore, rather than training the DNN model to tolerate refresh failures as done in EDEN~\cite{eden} and RANA~\cite{tu2018rana}, we opt to make alterations directly onto the DNN model to avoid refreshing long-lived data during forward and backward passes, and doing so without compromising the model's accuracy. 
DaDianNao~\cite{chen2014dadiannao} mitigates eDRAM refresh failures via bank splitting, which does not address data lifetime and retention concerns, therefore DaDianNao still requires on-demand eDRAM refresh with its associated power overheads.

In contrast, our work significantly reduces the eDRAM refresh by co-designing the DNN model (Section~\ref{sec:duplex-dnn-arch}), the computation scheduling (Section~\ref{sec:computation-pattern}) and the hardware accelerator (Section~\ref{sec:camel-hardware}), leading to a higher overall hardware efficiency.

\section{Duplex DNN for Fast On-Chip Learning}
\label{sec:camel-DuDNN}
While the reversibility feature discussed in Section~\ref{sec:reversible-dnn-intro} dramatically reduces memory requirements to store intermediate activations, this approach in turn imposes higher compute demands to recompute the intermediate values needed during the backward pass. Worse still, training a reversible DNN from scratch usually takes tens of thousands of iterations to converge, which further exacerbates the latency and energy consumption. A naive way to address this problem would be to reduce model size, but this degrades accuracy. Instead, we propose to judiciously train a subset of the model parameters to minimize training cost while maintaining accuracy.
\subsection{Duplex DNN Architecture}
\label{sec:duplex-dnn-arch}
\begin{figure}
    \centering
    \includegraphics[width=0.48\textwidth]{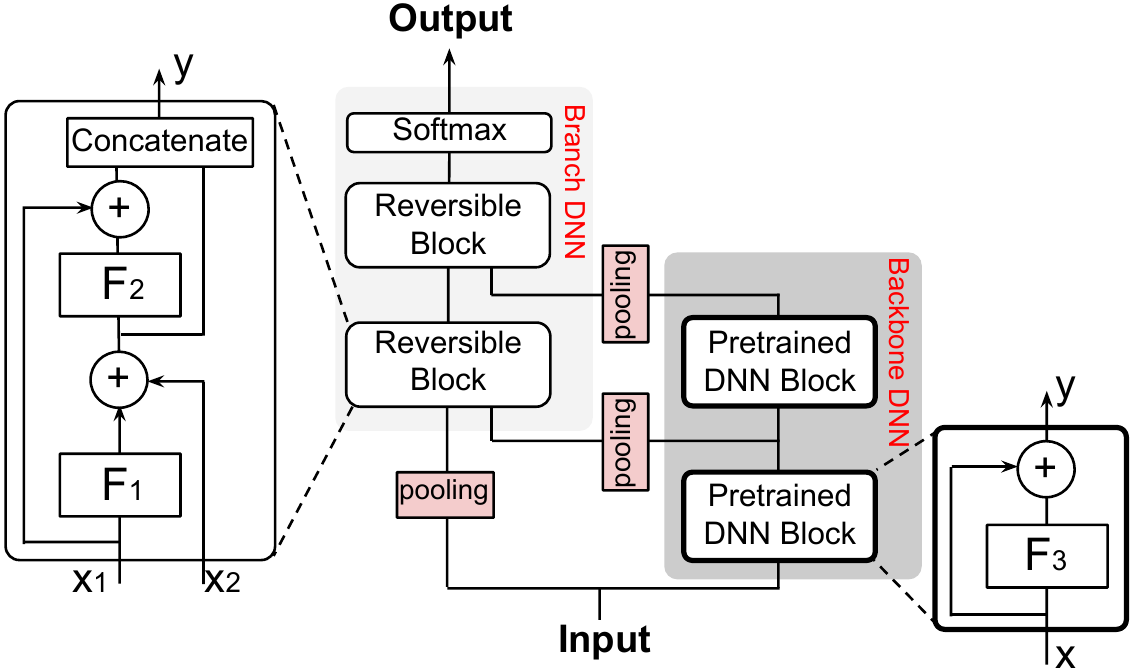}
    \caption{The architecture of DuDNN, which consists of a two-layer backbone DNN and a two-layer branch DNN that consists of reversible block shown in Figure~\ref{fig:rev-arch-ex}. For illustration simplicity, we concatenate outputs by making $y = [y_{1},y_{2}]$. We assume the backbone DNN uses residual architecture in this example, but other architectures can also be easily integrated.}
    \label{fig:duplex-dnn-ex}
    \squeezeup
\end{figure}
Figure~\ref{fig:duplex-dnn-ex} illustrates the proposed \textit{Duplex DNN (DuDNN)}, comprising a pre-trained, fixed \textit{Backbone DNN} (shown in dark grey) combined with a trainable \textit{Branch DNN} (shown in light grey). For efficient on-device training, the branch DNN comprises several reversible blocks, while the backbone DNN can be any DNN architecture, such as ResNet or Transformer. Moreover, the backbone DNN, whose size is larger than branch DNN, is pretrained offline using a large-scale training dataset (e.g., ImageNet). 

\subsection{Training and Inference Operation of Duplex DNN}
\label{sec:duplex-dnn-training}
\begin{figure}
    \centering
    \includegraphics[width=0.5\textwidth]{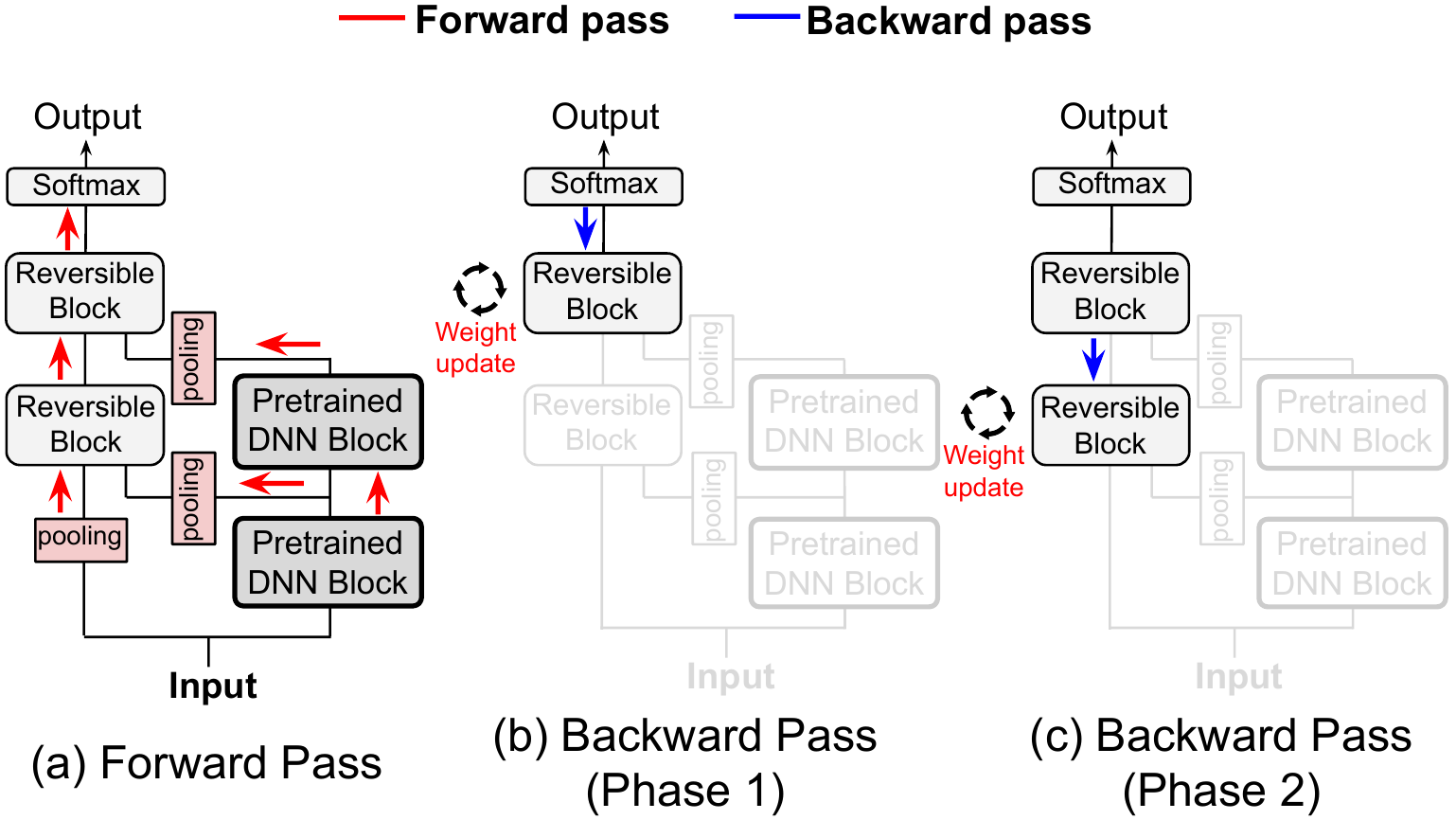}
    \caption{The training steps of two-layer DuDNN.}
    \label{fig:duplex-dnn-arch}
    \squeezeup
\end{figure}
The forward and backward passes of the DuDNN are described in Figure~\ref{fig:duplex-dnn-arch}. During the forward pass, the input sample is sent to both backbone and branch DNN paths (Figure~\ref{fig:duplex-dnn-arch}(a)). Each reversible block in the branch DNN also accepts an intermediate output from the backbone DNN block as an input, so that knowledge can be transferred from the backbone DNN to the branch DNN. During the backward pass, all the weights in the backbone DNN are frozen and only the parameters in the branch DNN are updated (Figure~\ref{fig:duplex-dnn-arch}(b,c)). The backbone DNN is designed to have a larger size than the branch DNN in order to better guide the learning of the branch DNN. Moreover, a larger backbone DNN will not cause a notable growth in compute load since the backbone DNN does not participate in the backward pass.

To further reduce training cost, we utilize the BFP quantization technique as described in Section~\ref{sec:bg:bfp}. This technique involves applying BFP representations to all the weights, activations, and gradients. During training, we divide each matrix operand into multiple groups and quantize them using the BFP format. However, it is important to note that the proposed DuDNN architecture is also compatible with other numerical types. 

We also add pooling operations along the backbone DNN connections to the branch DNN (red blocks in Figure~\ref{fig:duplex-dnn-ex}) as well as the raw input samples in order to relieve compute load. Since the amount of computations during the forward and backward passes both grow quadratically with input spatial size (i.e. $W\times H$), a smaller input translates to remarkable computational savings during branch DNN training. Evaluations in Section~\ref{sec:acc-ablation} show that aggressive pooling (e.g., pooling factor of 16) can be applied without impacting training accuracy. 

\subsection{Normalization Layers in Duplex DNN}
\label{sec:norm-ops}
Normalization layers (e.g., Batch Normalization (BN)~\cite{ioffe2015batch}) have been established as an important component in DNNs. Generally, normalization improves the convergence speed of the training process for DNNs across multiple fields including computer vision~\cite{szegedy2016rethinking,he2016deep} and natural language processing~\cite{vaswani2017attention,devlin2018bert}. 
Despite improving performance, the normalization operation exhibits large computation and storage overheads in both the forward and backward passes of training. 
The computation overhead is mainly due to the square root operation during standard deviation computations, while the storage overhead is caused by the fact that the normalization operation is not reversible. 
In other words, the input to the normalization layer during the forward pass must be saved for the backward pass operation. To improve hardware performance during training, we eliminate normalization layers in the branch DNN but keep the normalization layers in the backbone DNN.
Moreover, since the backbone DNN only participates in the forward pass, normalization parameters can be folded into the weights~\cite{ioffe2015batch} such that the associated computation overhead can be completely eliminated. Evaluation results in Section~\ref{sec:acc-ablation} show that even without the normalization layers, the branch DNN can still achieve high training accuracy and good convergence behavior under the guidance of the backbone DNN.

\section{Computational Pattern Design for Efficient eDRAM Performance}
\label{sec:computation-pattern}
There is an opportunity to carefully schedule the computations of DuDNN during both forward and backward passes to achieve optimal system performance and derive analytical expressions for the associated data lifetimes.

\begin{figure}
    \centering
    \includegraphics[width=0.5\textwidth]{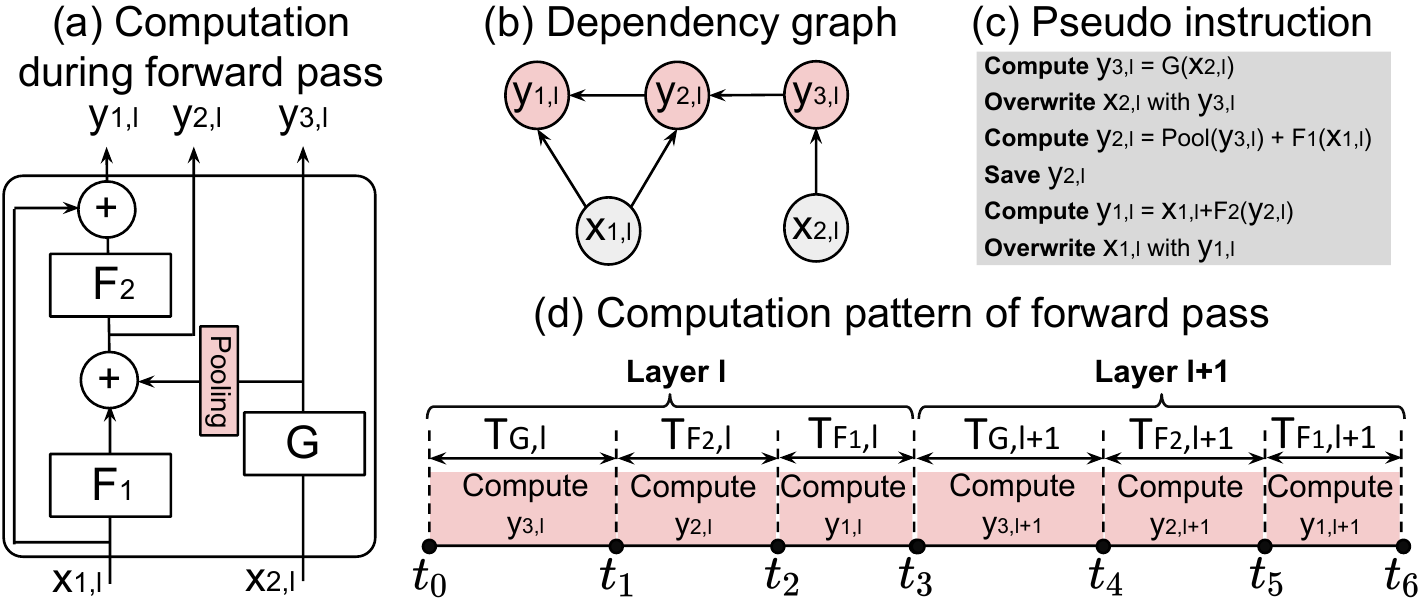}
    \caption{(a) DuDNN block architecture. (b) Dependency graph of forward pass. (c) The pseudo instruction to execute the forward pass. (d) The computation pattern of the forward pass with time.}
    \label{fig:forward-pass-scheduling}
    \squeezeup
\end{figure}
\subsection{Computation Scheduling During Forward Pass}
\label{sec:forward_pass_pattern}
We first discuss the computation pattern during the forward pass of DuDNN. Figure~\ref{fig:forward-pass-scheduling}(a) shows the architecture of a DuDNN block. The branch DNN at layer $l$ takes $x_{1,l}$ as the input and produces two outputs $y_{1,l}$ and $y_{2,l}$, meanwhile the backbone DNN takes $x_{2,l}$ as the input and generates $y_{3,l}$.

We first provide a numerical analysis on the data lifetime of convolutional-based architecture, assume $x_{1,l}$ has a shape of $B\times C^{in}_{F_{1},l}\times W_{F_{1},l}\times H_{F_{1},l}$ and $x_{2,l}$ has a shape of $B\times C^{in}_{G,l}\times W_{G,l}\times H_{G,l}$, where $B,C^{in}_{F_{1},l},C^{in}_{F_{2},l}$ and $C^{in}_{G,l}$ denote the batch size, the number of input channels in $F_{1}$, $F_{2}$ and $G$, respectively. $W_{F_{1},l}$, $H_{F_{1},l}$, $W_{F_{2},l}$, $H_{F_{2},l}$, $W_{G,l}$ and $H_{G,l}$ indicate the input spatial size (width and height) of $F_{1},F_{2}$ and $G$, respectively. Furthermore, define $C^{out}_{F_{1},l}$, $C^{out}_{F_{2},l}$, and $C^{out}_{G,l}$ as the number of filters in branch and backbone DNNs, and denote their weight filter kernel sizes to be $k_{F_{1},l}, k_{F_{2},l}$, and $k_{G,l}$, respectively. The processing latencies $T_{G,l}$, $T_{F_{1},l}$, and $T_{F_{2},l}$ of the CONV operations at $F_{1}$, $F_{2}$, and $G$ can be estimated as follows:
\begin{equation}
    \label{eqn:t1}
    T_{G,l} = \frac{N_{G,l}}{R} = \frac{BC^{in}_{G,l}W_{G,l}H_{G,l}k_{G,l}^{2}}{R}
\end{equation}
 \begin{equation}
    \label{eqn:t2}
    T_{F_{1},l} = \frac{N_{F_{1},l}}{R} = \frac{BC^{in}_{F_{1},l}W_{F_{1},l}H_{F_{1},l}k_{F_{1},l}^{2}}{R}
\end{equation}
 \begin{equation}
    \label{eqn:t3}
    T_{F_{2},l} = \frac{N_{F_{2},l}}{R} = \frac{BC^{in}_{F_{2},l}W_{F_{2},l}H_{F_{2},l}k_{F_{2},l}^{2}}{R}
\end{equation}
where $N_{G,l}$, $N_{F_{1},l}$, and $N_{F_{2},l}$ are the amount of computations associated with $G$, $F_{1}$, and $F_{2}$. $R$ is the hardware throughput.

Next, the dependency graph between input and output activations of a DuDNN block is shown in Figure~\ref{fig:forward-pass-scheduling}(b).
The dependency relations are indicated by the direction of the link in the dependency graph. For example, the directed link between $x_{2,l}$ and $y_{3,l}$ indicates $y_{3,l}$ is generated using $x_{2,l}$. 
The dependency relationships shown in Figure~\ref{fig:forward-pass-scheduling}(b) translate into the computation pattern of the forward pass illustrated in Figure~\ref{fig:forward-pass-scheduling}(d). And, during the computation of the forward pass, we overwrite any activations that will not be used by the future operations. For example, once $y_{3,l}$ is produced, $x_{2}$ can be overwritten by $y_{3,l}$ since $x_{2,l}$ will not be used any more. This leads to the pseudo instructions described in Figure~\ref{fig:forward-pass-scheduling}(c).

\begin{figure}
    \centering
    \includegraphics[width=0.5\textwidth]{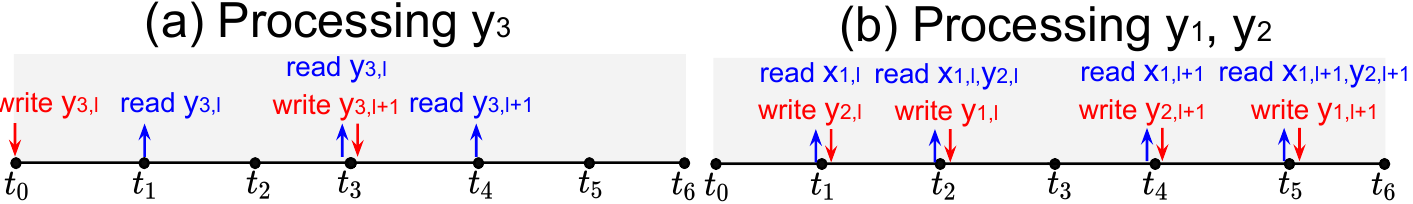}
    \caption{The memory access patterns during the forward pass of layer l and l+1.}
    \label{fig:forward-pass-datalife}
    \squeezeup
\end{figure}

The memory access patterns to compute output activations $y_{1}, y_{2}, y_{3}$ at layer l and l+1 are shown in Figure~\ref{fig:forward-pass-datalife}. The data lifetime corresponds to the maximum gap between a memory read operation (blue arrow in Figure~\ref{fig:forward-pass-datalife}) and the most recent memory write operation (red arrow in Figure~\ref{fig:forward-pass-datalife}) prior to this read operation. 
Specifically, the maximum data lifetime $T_{f}$ for the forward pass can be expressed as follows:
\begin{equation}
    \label{eqn:forward_pass_datalife}
    T_{f} = \underset{1 \leq l \leq L}{\max} (T_{y_{1},l}^{f},T_{y_{2},l}^{f},T_{y_{3},l}^{f})
\end{equation}
where $L$ is the total number of layers, $T_{y_{1},l}^{f},T_{y_{2},l}^{f},T_{y_{3},l}^{f}$ are the maximum data lifetime in computing $y_{1}, y_{2}$ and $y_{3}$ at layer l and l+1. Based on Figure~\ref{fig:forward-pass-datalife}, we have $T_{y_{3},l}^{f} = t_{3}-t_{0} = T_{G,l} + T_{F_{1},l} + T_{F_{2},l}$, $T_{y_{1},l}^{f} = t_{5}-t_{2} = T_{F_{1},l} + T_{G,l+1} + T_{F_{2},l+1}$ and $T_{y_{2},l}^{f} = t_{5}-t_{1} = T_{F_{1},l} + T_{F_{2},l} + T_{G,l+1} + T_{F_{2},l+1}$. We assume the pooling, ReLU, and residual addition have much lower processing times than the CONV operation. Given the layer configuration and hardware throughput, we can calculate the maximum data lifetime and the required eDRAM retention time by substituting Equation~\ref{eqn:t1}, Equation~\ref{eqn:t2}, and Equation~\ref{eqn:t3} onto Equation~\ref{eqn:forward_pass_datalife}, resulting in a closed-formed analytical solution for the data lifetime for the forward pass.
\subsection{Computation Scheduling for Backward Pass}
\label{sec:backward_pass_pattern}
\begin{figure}
    \centering
    \includegraphics[width=0.48\textwidth]{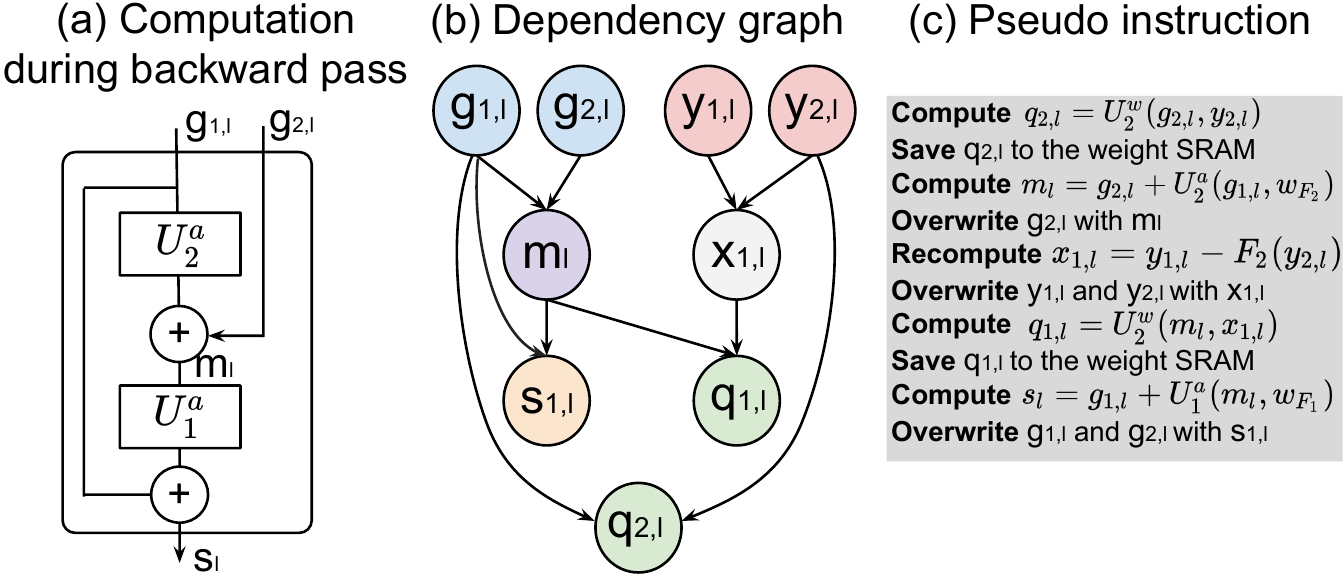}
    \caption{(a) The computation of the input gradient. (b) Dependency graph of backward pass. (c) The pseudo instruction to execute backward pass. }
    \label{fig:backward-pass-scheduling}
    \squeezeup
\end{figure}

To accommodate the computing limitations of edge devices effectively, we restrict the execution of backward operations exclusively within the branch DNN. This approach allows the branch DNN to efficiently adapt to user data while minimizing training costs. Given the output activations $y_{1,l}, y_{2,l}$ and the corresponding output gradients $g_{1,l}, g_{1,l}$, they are processed to generate the input gradient $s_{l}$ and weight gradients $q_{1,l}$ and $q_{2,l}$ for $w_{F_{1}}$ and $w_{F_{2}}$, where $w_{F_{1}}$ and $w_{F_{2}}$ denote the DNN weights presented in $F_{1}$ and $F_{2}$, respectively. Figure~\ref{fig:backward-pass-scheduling}(a) shows the computations during the backward pass to produce the input gradient $s_{l}$ with the output gradients $g_{1,l}$ and $g_{2,l}$. $U^{a}_{1}$ and $U^{a}_{2}$ represent the functions for input gradient computations at $F_{1}$ and $F_{2}$, respectively. Given output gradients $g_{1,l}, g_{2,l}$, the intermediate result $m_{l} = g_{2,l} + U_{2}^{a}(g_{1,l},w_{F_{2}})$ is first computed. Then, the input gradient $s_{l}$ is computed with $s_{l}=g_{1,l} + U_{1}^{a}(m_{l},w_{F_{1}})$.

Similar to the forward pass, we can find the required data lifetime during the backward pass. For convolutional-based architecture, assume all the output activations $y_{1,l}, y_{2,l}$ and output gradients $g_{1,l}, g_{2,l}$ have a shape of $B\times C^{out}_{F_{2},l}\times W_{F_{2},l}\times H_{F_{2},l}$. The latencies of the input activation, weight gradient, and input gradient computation at $F_{1}$, $F_{2}$ can be computed as follows:
\begin{equation}
    \label{eqn:t4}
    T_{U_{2}^{a},l} = T_{U_{2}^{w},l} = T_{F_{2},l} = \frac{BC^{out}_{F_{2},l}W_{F_{2},l}H_{F_{2},l}k_{F_{2},l}^{2}}{R}
\end{equation}
 \begin{equation}
    \label{eqn:t5}
     T_{U_{1}^{a},l} = T_{U_{1}^{w},l} = T_{F_{1},l} = \frac{BC^{out}_{F_{1},l}W_{F_{1},l}H_{F_{1},l}k_{F_{1},l}^{2}}{R}
\end{equation}
where $T_{U_{1}^{a},l}, T_{U_{1}^{w},l},  T_{F_{1},l}$ and $T_{U_{2}^{a},l}, T_{U_{2}^{w},l}, T_{F_{2},l}$ represent the processing latencies for input gradient, weight gradient, and input activation computations at $F_{1}$ and $F_{2}$, respectively. 

\begin{figure}
    \centering
    \includegraphics[width=0.48\textwidth]{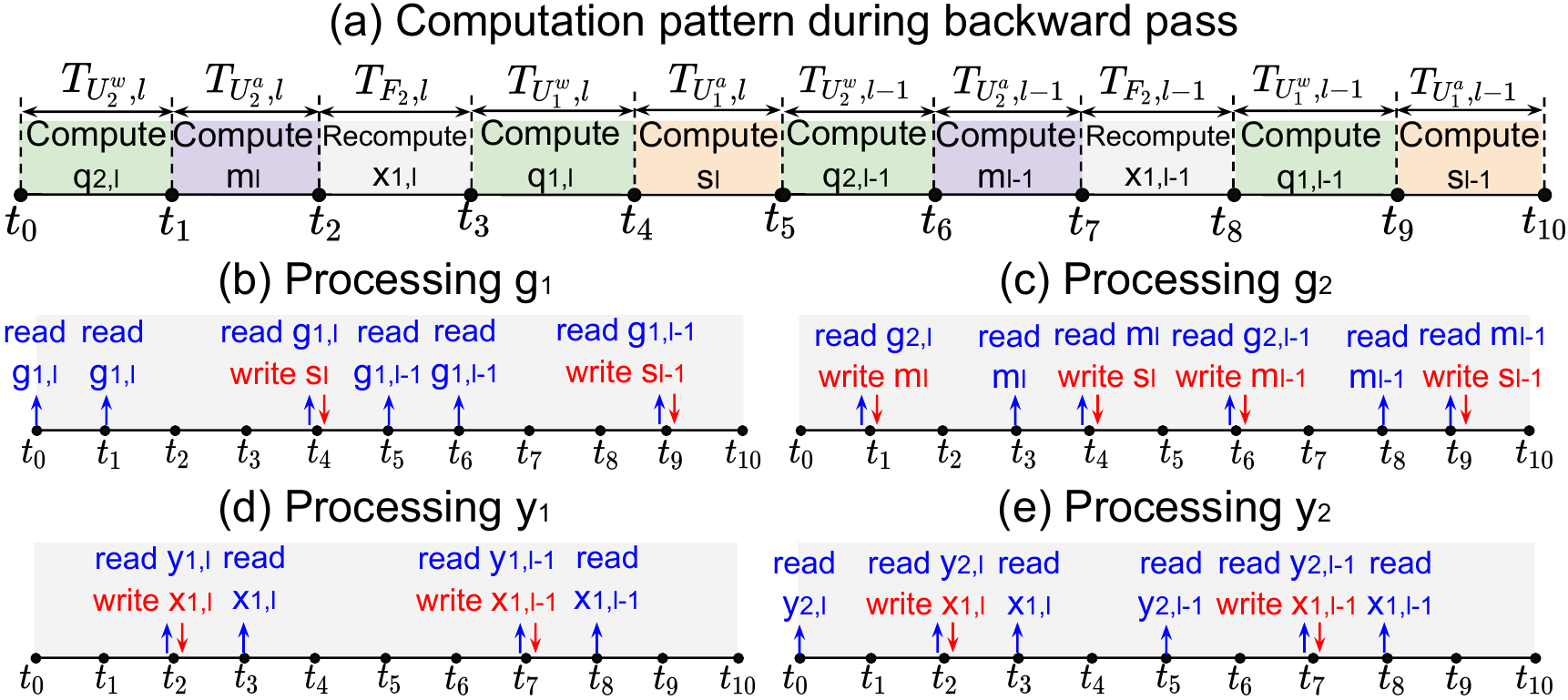}
    \caption{(a) The computation pattern of backward pass. (b-e) The data access patterns at the memory banks during the backward pass of layer l and l-1.}
    \label{fig:backward-pass-datalife}
\end{figure}

The dependency graph for the backward pass is given in Figure~\ref{fig:backward-pass-scheduling} (b), and Figure~\ref{fig:backward-pass-datalife}(a) describes the optimal computation pattern with the least memory consumption and shortest data lifetime, where any intermediate results that will not be used by future operations are overwritten. The instructions associated with the computation pattern are shown in Figure~\ref{fig:backward-pass-scheduling}(c).
With the aforementioned computation pattern, the memory access patterns during the processing of the output gradients $g_{1},g_{2}$ and output activations $y_{1},y_{2}$ across layer l and layer l-1 are shown in Figure~\ref{fig:backward-pass-datalife}(b)-(e).
According to this pattern, the maximum data lifetime $T_{b}$ during the backward pass are computed as follows:
\begin{equation}
    \label{eqn:backward_pass_datalife}
    T_{b} = \underset{1 \leq l \leq L}{\max} (T_{g_{1},l}^{b},T_{g_{2},l}^{b},T_{y_{1},l}^{b},T_{y_{2},l}^{b})
\end{equation} 
$T_{g_{1},l}^{b},T_{g_{1},l}^{b},T_{y_{1},l}^{b},T_{y_{2},l}^{b}$ are the maximum data lifetime when processing $g_{1}, g_{2}, y_{1}$ and $y_{2}$ during the backward pass of layer l and l-1. As shown in Figure~\ref{fig:backward-pass-datalife}, we have $T_{g_{1},l}^{b} = t_{9}-t_{4} = T_{U_{1}^{a},l} +T_{U_{2}^{w},l-1} + T_{U_{2}^{a},l-1} + T_{F_{2},l-1}+T_{U_{1}^{w},l-1}$, $T_{g_{2},l}^{b} = t_{4}-t_{1} = T_{U_{2}^{a},l} + T_{F_{2},l} + T_{U_{1}^{w},l}$ and $T_{y_{1},l}^{b} = T_{y_{2},l}^{b}= t_{7}-t_{2} = T_{F_{2},l}+T_{U_{1}^{w},l} + T_{U_{1}^{a},l} +T_{U_{2}^{w},l-1} + T_{U_{2}^{a},l-1}$. 
We can calculate the maximum data lifetime by substituting Equation~\ref{eqn:t4} and Equation~\ref{eqn:t5} onto Equation~\ref{eqn:backward_pass_datalife}. 

Finally, the maximum data lifetime during DNN training can be expressed as follows:
\begin{equation}
    \label{eqn:total_datalife}
    T_{data} = max(T_{f},T_{b})
\end{equation}
where $T_{f}$ and $T_{b}$ are the maximum data lifetimes during the forward and backward pass as defined in Equation~\ref{eqn:forward_pass_datalife} and Equation~\ref{eqn:backward_pass_datalife}.

\section{CAMEL Hardware Architecture} 
\label{sec:camel-hardware}
\begin{figure}
    \centering
    \includegraphics[width=0.48\textwidth]{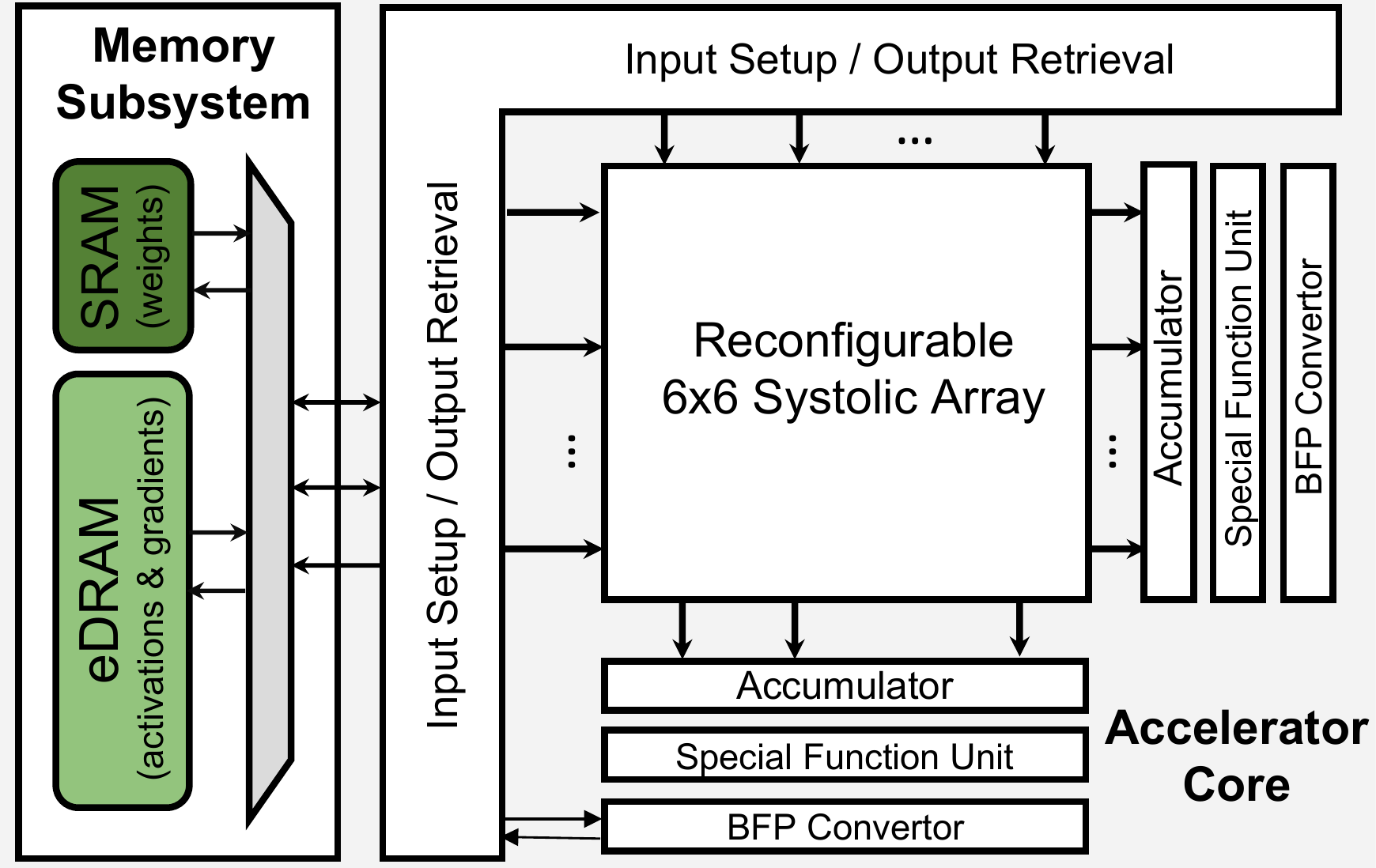}
    \caption{The CAMEL training accelerator system.}
    \label{fig:camel_accel}
    \squeezeup
\end{figure}

To promote greater on-chip training performance, we propose a reconfigurable hardware accelerator that primarily comprises a systolic array core and a hybrid eDRAM-SRAM memory subsystem (Figure~\ref{fig:camel_accel}). 

\subsection{Systolic Array Core} 
The systolic array core consists of a 2D 6-by-6 bidirectional processing element (PE) array that receives staggered inputs from the \textit{I/O} controller and then drains out the computed partial sums onto a post-processing unit composed of accumulator, special function, and BFP converter units. 
The accelerator benefits from the higher computational accuracy and efficiency of BFP encapsulated inside the PE array and the greater hardware density of the fixed-point based post-processing unit. 
The accumulator aggregates the partial sums into a register file with 64 entries. The special function unit (SFU) computes, in a vectorized fashion, several mathematical and non-linear functions such as ReLU, elementwise addition, softmax, and square root, all of which get invoked during the training passes. The fixed-point outputs of the SFU get converted back into the BFP datatype in order to be stored inside the memory subsystem. 

\begin{figure}
    \centering
    \includegraphics[width=0.48\textwidth]{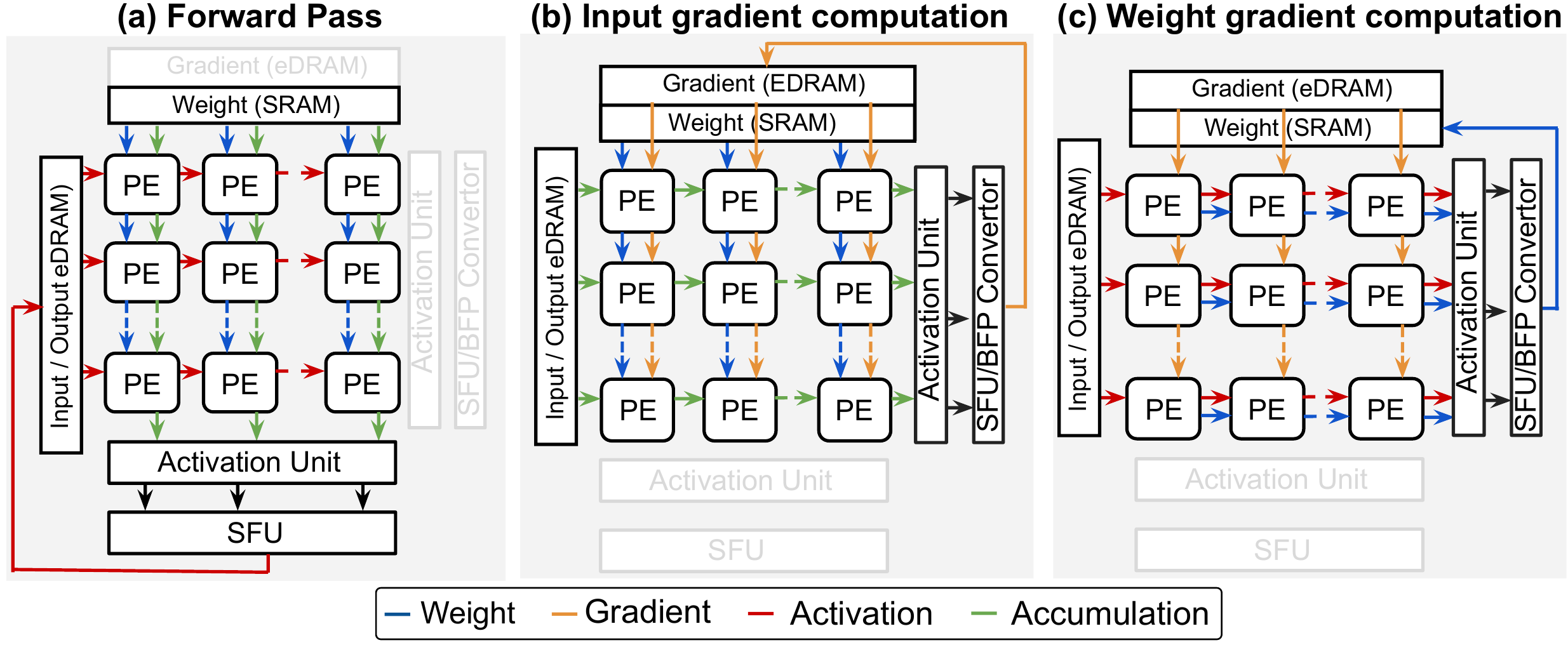}
    \caption{Accelerator workflows during the computation of the training passes.}
    \label{fig:accel_dataflows}
    \squeezeup
\end{figure}

The systolic array core can be reconfigured to operate in three operational workflows to support the transposed matrix multiplication~\cite{zhang2022fast} as illustrated in Figure~\ref{fig:accel_dataflows}.
During the forward and backward propagation for the input gradient, the accelerator adopts a weight-stationary dataflow wherein the weights are first pre-loaded from the weight SRAM into the PE weight register and then the inputs are streamed into the array in a staggered cadence. During the forward propagation, activations are fed to the PE array from left to right and the partial sums are accumulated from top to bottom (Figure~\ref{fig:accel_dataflows}(a)). In contrast, the backward propagation for the input gradient operates in the reverse direction, i.e., the gradient data is streamed from top to bottom while the results are accumulated from left to right (Figure~\ref{fig:accel_dataflows}(b)). 
This structure ensures matrix-matrix multiplications efficiently operate on the correct dimensionality of the different input operands invoked during the forward and backward passes.
Furthermore, the systolic array is configured with an accumulation-stationary dataflow during the backward propagation for the weight gradient. In this mode, the \textit{IO} control unit feeds in the activations and gradients simultaneously into the array and the result accumulates inside the PE until weight gradient tiles are fully computed. At this point, the PEs drain out the accumulated results from left to right for post-processing and weight updating in SFU, then the updated weights are stored in the weight SRAM (Figure~\ref{fig:accel_dataflows}(c)). 

\begin{figure}
    \centering
    \includegraphics[width=0.48\textwidth]{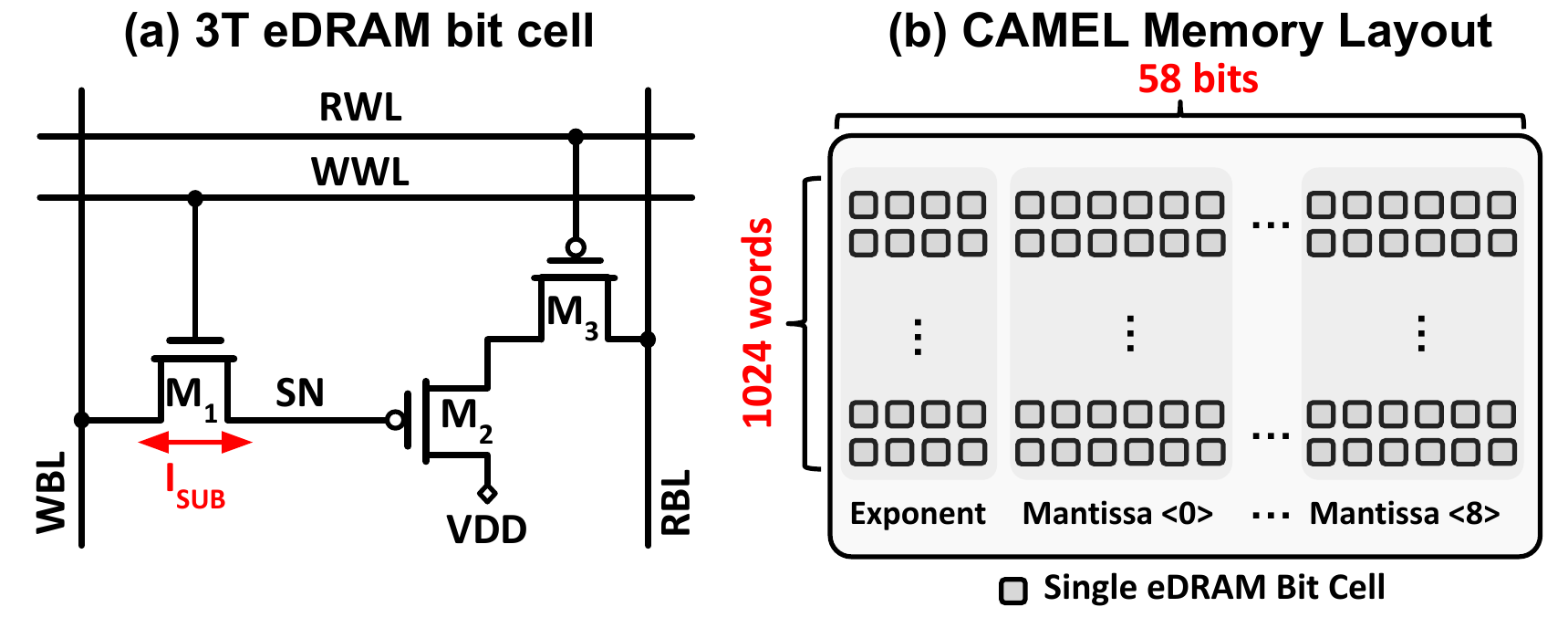}
    \caption{(a) Schematic view of a 3T eDRAM bit cell, highlighting the main charge/discharge path ($I_{\textrm{SUB}}$) for the storage node (SN). RWL -- read word line, RBL -- read bit line, WWL -- write word line, and WBL -- write bit line. (b) Block floating point storage mapping within a $58 \times$ 1024 eDRAM array.}
    \label{fig:edram_map}
    \squeezeup
\end{figure}
\begin{figure}
    \centering
    \includegraphics[width=0.48\textwidth]{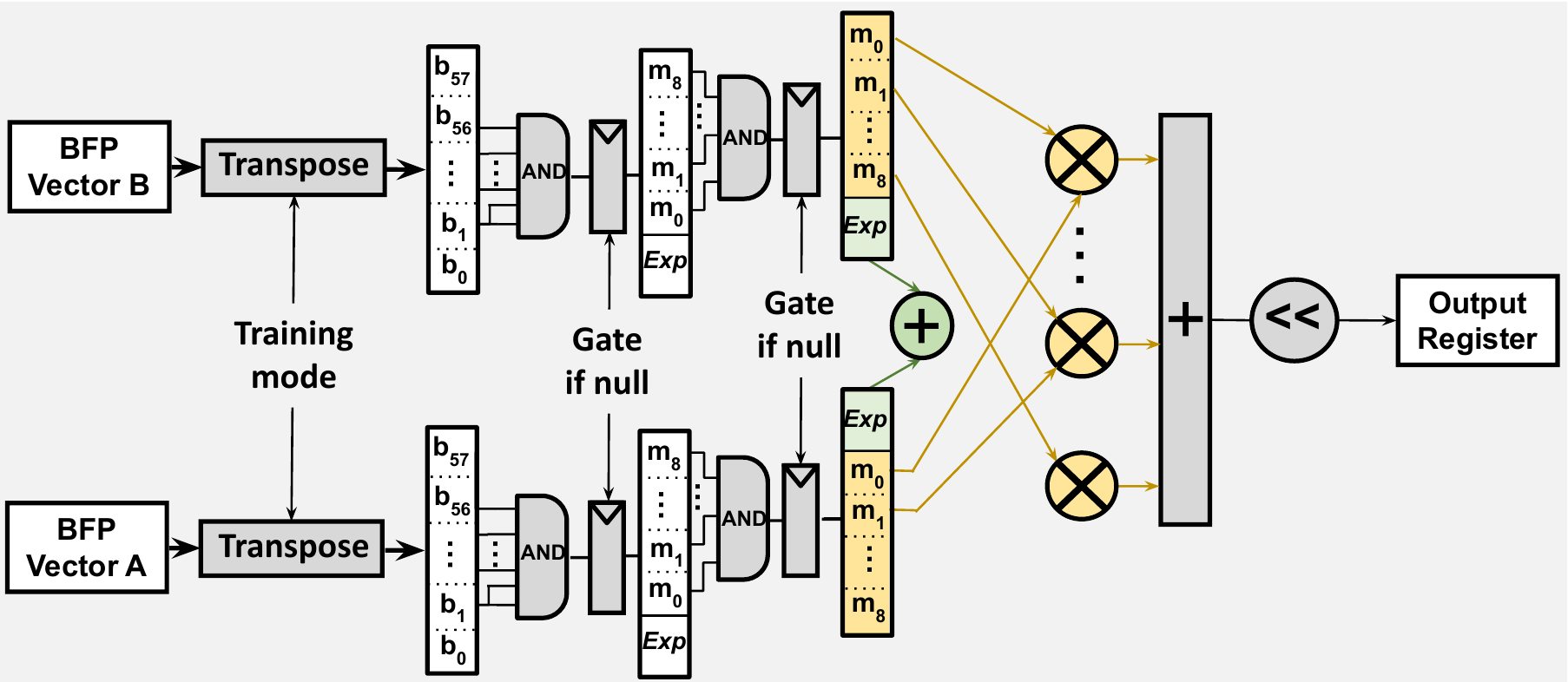}
    \caption{Architecture of the CAMEL processing element.}
    \label{fig:pe}
    \squeezeup
\end{figure}
\subsection{Processing Element Design}
\begin{figure}
    \centering
    \includegraphics[width=0.48\textwidth]{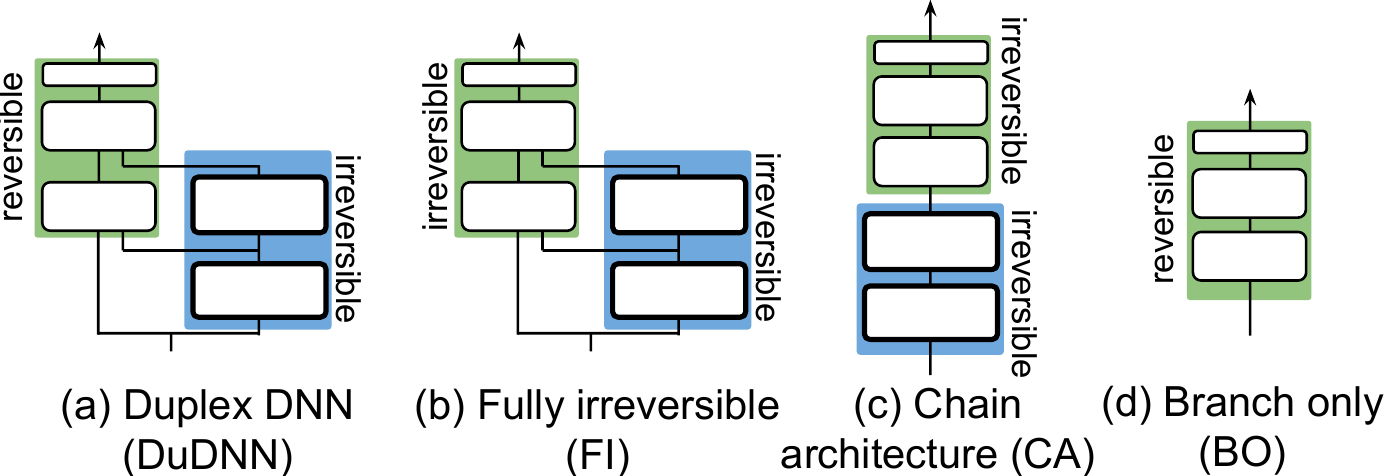}
    \caption{Baseline algorithms for accuracy evaluation. For each baseline architecture, its trainable part is highlighted in green while the frozen part is highlighted in blue.}
    \label{fig:baselines}
    \squeezeup
\end{figure}
The architecture of the processing element (PE) of the systolic array is shown in Figure~\ref{fig:pe}. The PE receives two tensor operands and performs multiply-and-accumulate (MAC) computations. In this work, we 
apply BFP quantization technique on all the weights, activations, and gradients. However, it is again worth noting that the proposed PE design can be adapted to accommodate various data formats as needed.

Depending on the training mode (i.e., forward or backward pass), the BFP vector gets transposed in-place.
To promote greater energy efficiency, the PE is equipped with two gating checkpoints before the MAC hardware resources are utilized. First, in case one of the two BFP operands are zero, the PE immediately skips to write zero onto the output register. Otherwise, another check is performed on the mantissa bits whereby the mantissa multiplier is gated in case all the mantissa bits in one of the two operands are zero. The BFP MAC executes a \textit{left-shift} operation on the product-sum of the mantissas by the sum of the exponents. 

\subsection{Hybrid eDRAM-SRAM Memory Subsystem} 
\label{sec:mem-layout}
The memory subsystem collects and unifies, across channels and filters, the input/output activations and gradients, as well as the master weights. 
Given the ultimate goal of the training regime is to learn more accurate representations of the master weights, which potentially will be used in subsequent inference tasks, the learned weights are then stored in a static medium (i.e., SRAMs). On the other hand, the transient activation and gradient parameters, which represent the majority of the total training data (Figure~\ref{fig:dnn-training-ex}(b)), are housed in eDRAM memories. 
The hybrid SRAM-eDRAM memory partitioning provides the benefits of high density eDRAM storage for the majority of training data and stable long-term SRAM storage for the weights that constitute the final solution of the on-chip learning task. 

In CAMEL, we utilize a BFP format featuring a 4-bit shared exponent field, 5-bit mantissa, and 1-bit sign for each of the nine numbers within the BFP group. As a result, every BFP number group consists of a total of $4 + (5 + 1) \times 9 = 58$ bits, leading to an effective precision of 6.4 bits per number for weights, activations, and gradients. In our experiments, we demonstrate that this format can effectively reduce the implementation cost without compromising overall model accuracy. To accommodate all this data, we integrate twelve 48KB eDRAM banks and six 8KB SRAM banks into the memory subsystem.

\subsection{eDRAM Design}

We designed the eDRAM architecture for the CAMEL system based on the architectures proposed in~\cite{giterman20201, narinx2019edram}, which have been demonstrated effective across various transistor technologies such as FinFET. With the same capacity, a 3T eDRAM design allows for more than $2\times$ reduction on memory footprint compared with SRAM. However, eDRAM suffers comparatively short retention time and, therefore, may consume additional energy to refresh the stored data.

As Figure~\ref{fig:edram_map}(a) shows, the main leakage path for the storage node (SN) to charge/discharge is I$_{\textrm{SUB}}$, passing through the write transistor M$_{\textrm{1}}$. Therefore, to extend the retention time of the memory cell, we employ well studied leakage mitigation techniques such as write wordline (WWL) over- and under-drive \cite{chun2011edram, giterman20201}. Through an extended retention time and by co-designing our DNN training algorithm with the model architecture and hardware to support it, we completely eliminate the need for explicit eDRAM refresh. Figure~\ref{fig:edram_map}(b) illustrates the 3T eDRAM array layout for CAMEL. Here, we use memory arrays of 58 bits and 1024 words to better match the BFP data shape of our DNN algorithm. In other words, the 58 bits corresponds to one 4-bit shared exponent stored along with nine 6-bit signed mantissas.

\section{Accuracy Evaluation}
\label{sec:accu-evaluation}
In this section, we first evaluate the model accuracy of DuDNN across different DNN architectures and datasets. Additionally, we conduct an ablation study to explore the behavior of DuDNN under various settings. 

\subsection{Settings for Duplex Architecture Training}
\label{sec:acc-eval}
We adopt different architectures for the backbone DNN, including CNNs (ResNet-18, ResNet-34, ResNet-50~\cite{he2016deep}, VGG-16~\cite{simonyan2014very}) and vision transformer (ViT)~\cite{dosovitskiy2020image}. The backbone DNN is pre-trained offline with the CIFAR-100~\cite{krizhevsky2014cifar} or ImageNet (ILSVRC 2012)~\cite{deng2009imagenet} datasets. The pre-trained backbone DNN is then used to guide the training of branch DNN over CIFAR-10~\cite{krizhevsky2014cifar} or Tiny-ImageNet~\cite{tinyimagenet} datasets.

To fairly evaluate the model accuracy of DuDNN, we compare to other baseline architectures that are shown in Figure~\ref{fig:baselines}. The \textit{fully irreversible (FI)} architecture applies the irreversible architecture (e.g., residual architecture) to both backbone and branch DNNs. The purpose of this baseline is to evaluate the impact of reversible architectures on validation accuracy. The \textit{Chain architecture (CA)} concatenates the reversible branch DNN with the pre-trained backbone DNN. During the training process, only the reversible components are learnable. The \textit{branch only (BO)} architecture trains the branch DNN solely, without the support from the backbone DNN. BO is used to evaluate the impact of the backbone DNN on the convergence behavior of branch DNN. For fair comparisons, every layer within the learnable components of DuDNN, FI, CA, and BO, maintains an equal number of learnable parameters.

\subsection{Accuracy Evaluation on Different Baselines}
\label{sec:cnn-evaluation}

\begin{table*}
 \centering
 \caption{The training accuracies of each method. The numbers outside the brackets represent the BFP training accuracy.}
\begin{adjustbox}{width=2\columnwidth,center}
    \begin{tabular}{ccccccccccc}\toprule 
    & \multirow{2}{*}{\begin{tabular}[c]{@{}c@{}}Architectures \end{tabular} }
    & \multicolumn{2}{c}{DuDNN} & \multicolumn{2}{c}{FI} & \multicolumn{2}{c}{CA} & \multicolumn{2}{c}{BO} \\
    \cmidrule(lr){3-4} \cmidrule(lr){5-6} \cmidrule(lr){7-8} \cmidrule(lr){9-10}
    &  & CIFAR-10  & Tiny-ImageNet  & CIFAR-10  & Tiny-ImageNet  & CIFAR-10  & Tiny-ImageNet  & CIFAR-10  & Tiny-ImageNet \\ \toprule 
    & Branch-4 + ResNet-18   & 87.08\% & 62.12\%   & 87.20\%  & 62.56\%  & 78.33\% & 58.80\%  & 55.32\%  & 34.30\% \\
    & Branch-5 + ResNet-34   & 88.81\%  & 64.82\%    & 88.83\%   & 65.39\%  & 79.89\%  & 59.49\%   & 57.80\%   & 36.55\%  \\
    & Branch-6 + ResNet-50   & 89.22\%  & 65.67\%    & 89.33\%   & 66.34\%   & 81.14\%  & 61.06\%   & 61.62\% & 38.56\% \\
    & Branch-6 + ViT-12      & 89.03\%  & 65.97\%  & 89.11\% & 66.30\%  & 80.98\% & 61.42\%   & 60.98\%   & 38.60\%  \\
    & Branch-6 + VGG-16      & 89.15\% & 66.30\%   & 89.13\%  & 66.78\% & 80.99\%  & 62.05\% & 61.60\%   & 38.49\%  \\
    & Branch-8 + ViT-16      & 91.45\% & 66.76\%   & 91.59\%  & 66.93\%  & 81.83\% & 62.88\%  & 62.18\%   & 40.42\% \\
    & Branch-12 + ResNet-50  & 92.03\%  & 68.48\%   & 92.12\%  & 68.72\%  & 82.24\% & 64.11\%  & 63.45\%  & 42.66\% \\
    & Branch-12 + ViT-16     & 92.14\%  & 68.20\%   & 92.28\% & 68.59\%  & 82.08\%  & 63.97\% & 63.26\% & 42.88\% \\
\midrule
    \end{tabular}
\end{adjustbox}
\label{tal:acc-eval}
\end{table*}

We compare DuDNN performance versus different baseline architectures across different datasets. We apply BFP quantization described in Section~\ref{sec:mem-layout} for training all the DNNs. For the backbone, which contains a CNN, they are pre-trained with CIFAR-100 or ImageNet with mini-batch of size 512 for 90 epochs. We apply the SGD optimizer with an initial learning rate of $10^{-3}$, dividing it by 10 at 30 and 60 epochs. The weight decay and momentum are set to 0.0005 and 0.9, respectively.
For ViT, we adopt a pre-trained 12-layer model (ViT-12) and 16-layer model (ViT-16) with 12 heads and a hidden size of 768 from the Model Zoo website \cite{vitofficial}. We finetune the models with the BFP format on ImageNet for 60 epochs, which will then be used as the backbone DNN to guide the branch DNN training.
The branch DNN associated with the ViT-based backbone consists of multiple reversible blocks, each reversible block contains two attention modules. We train the learnable components in DuDNN, FI, CA, and BO with a mini-batch size of 64 for 20 epochs. For CNN-base DuDNN, we adjust the pooling layers such that the transferred knowledge from the backbone to branch DNN has a spatial size of $7 \times 7$ or smaller. We also simulate readout error for the eDRAM, where 99.9\% of the stored values are read correctly, while the remaining 0.1\% are read out as random noise.


Table~\ref{tal:acc-eval} presents the training accuracies of various DNNs. For BO, the branch DNN is trained from scratch without the backbone. We make the following observations. DuDNN achieves comparable accuracy as FI, which adopts an irreversible architecture in their branch DNN. However, compared with DuDNN, FI requires a tremendous amount of storage to buffer the intermediate activations and further incurs a large overall memory power and latency cost, as indicated in Section~\ref{sec:hardware-perf}. DuDNN outperforms CA because it leverages more intermediate outputs from the backbone DNN, providing it with a wealth of supporting guidance. On the other hand, CA only uses the output from the backbone DNN, which is less informative compared to DuDNN. Lastly, BO highlights the significance of the backbone DNN since without it, the branch DNN results in significantly lower accuracy. 

We also conducted evaluations of DuDNN training using the GLUE dataset~\cite{gluedataset}, which encompasses multiple text classification tasks. For these experiments, we utilized a pretrained 12-layer BERT model~\cite{devlin2018bert} as the backbone DNN and fine-tuned a six-layer Branch DNN. Each reversible block in the Branch DNN follows an attention-based architecture similar to the BERT block. All DNNs were trained with 8 epochs using BFP precision. The results presented in Table~\ref{tab:text-result} demonstrate that DuDNN outperforms CA and achieves comparable performance to FI across all datasets.

\subsection{Ablation Studies on DuDNN Accuracy}
\label{sec:acc-ablation}
\begin{figure}
    \centering
    \includegraphics[width=0.5\textwidth]{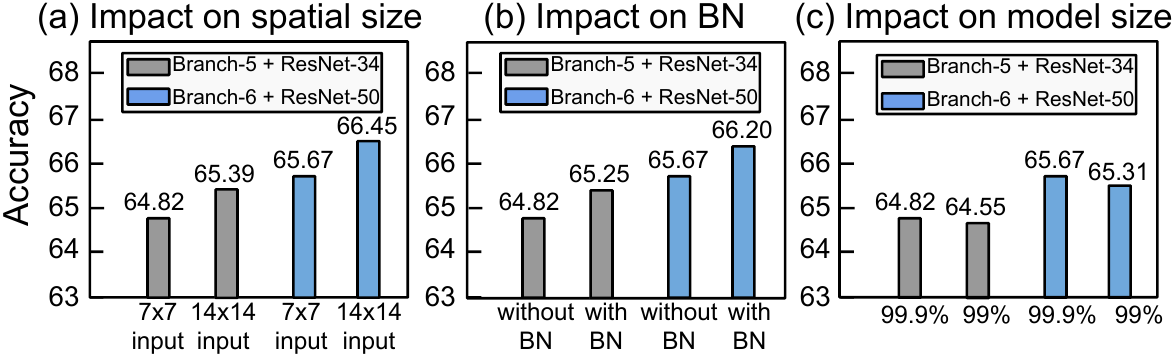}
    \caption{Ablation studies on DuDNN.}
    \label{fig:ablation}
    \squeezeup
\end{figure}

Next, we investigate the impact of the pooling layer and normalization layer on DuDNN accuracy. Specifically, we adjust the pooling layer setting to increase the spatial size of the transferred information from $7 \times 7$ to $14 \times 14$, and evaluate the changes on validation accuracies across different DNN architectures. Figure~\ref{fig:ablation}(a) shows the DuDNN performances under a spatial size of $7 \times 7$ and $14 \times 14$ for Branch-5+ResNet-34 and Branch-6+ResNet-50 on Tiny-ImageNet. We notice that a spatial size of $14 \times 14$ does not provide too much improvement on accuracy. Worse still, compared with a spatial size $7 \times 7$ information, a spatial size of $14 \times 14$ will lead to a $4\times$ increase on computation.

Figure~\ref{fig:ablation}(b) illustrates the impact of the normalization layer on DuDNN accuracy. We add a normalization layer in each layer of Branch DNN in DuDNN and evaluate the change in accuracy. We observe that introducing normalization within the branch DNN does not offer an obvious improvement in accuracy. A similar trend hold for other DNN architectures.

Lastly, in Figure~\ref{fig:ablation}(c), we examine the influence of eDRAM readout error on DuDNN accuracy. Specifically, we assess DuDNN's performance by trading off the percentage of correct eDRAM reads from 99.9\% to 99\% for longer retention time, while keeping all other training settings unchanged. The results reveal that the impact on DuDNN accuracy is small, being less than 0.4\% for both Branch-5+ResNet-34 and Branch-6+ResNet-50 models. We observe a similar trend for the other DNN architectures as well.


\section{Hardware Evaluation}
\label{sec:hw-evaluation}
In this section, we provide a comprehensive evaluation of the CAMEL system when running various workloads specified in the Section~\ref{sec:accu-evaluation}.
\subsection{Hardware Settings of CAMEL System}
\label{sec:hw-setting}

We start by evaluating hardware performance of the CAMEL system described in Section~\ref{sec:camel-hardware}. The CAMEL accelerator is designed in synthesizable SystemC and its verilog RTL is generated by the Catapult High-Level Synthesis (HLS) tool~\cite{catapult}. HLS directives are uniformly set with the goal to maximize throughput on the pipelined design. During the HLS process, the SRAM and eDRAM arrays are mapped to bit-accurate verilog templates. The following evaluation results are measured at 500MHz using a commercial 16nm process node. The CAMEL system contains a $6 \times 6$ systolic array, where each systolic cell can perform a dot product between two groups of nine BFP numbers within one cycle. All on-chip intermediate activations and gradients are stored in the eDRAM, while the weights are saved in the SRAM. A 4GB DRAM, as specified in~\cite{dramconfig}, is used as the second tier buffer. We implemented the CAMEL compiler in Python. The compiler produces hardware instructions and control configurations based on the computation pattern described in Section~\ref{sec:computation-pattern}. 

\begin{figure}
\centering
\begin{minipage}{0.23\textwidth}
    \centering
    \includegraphics[width=0.98\textwidth]{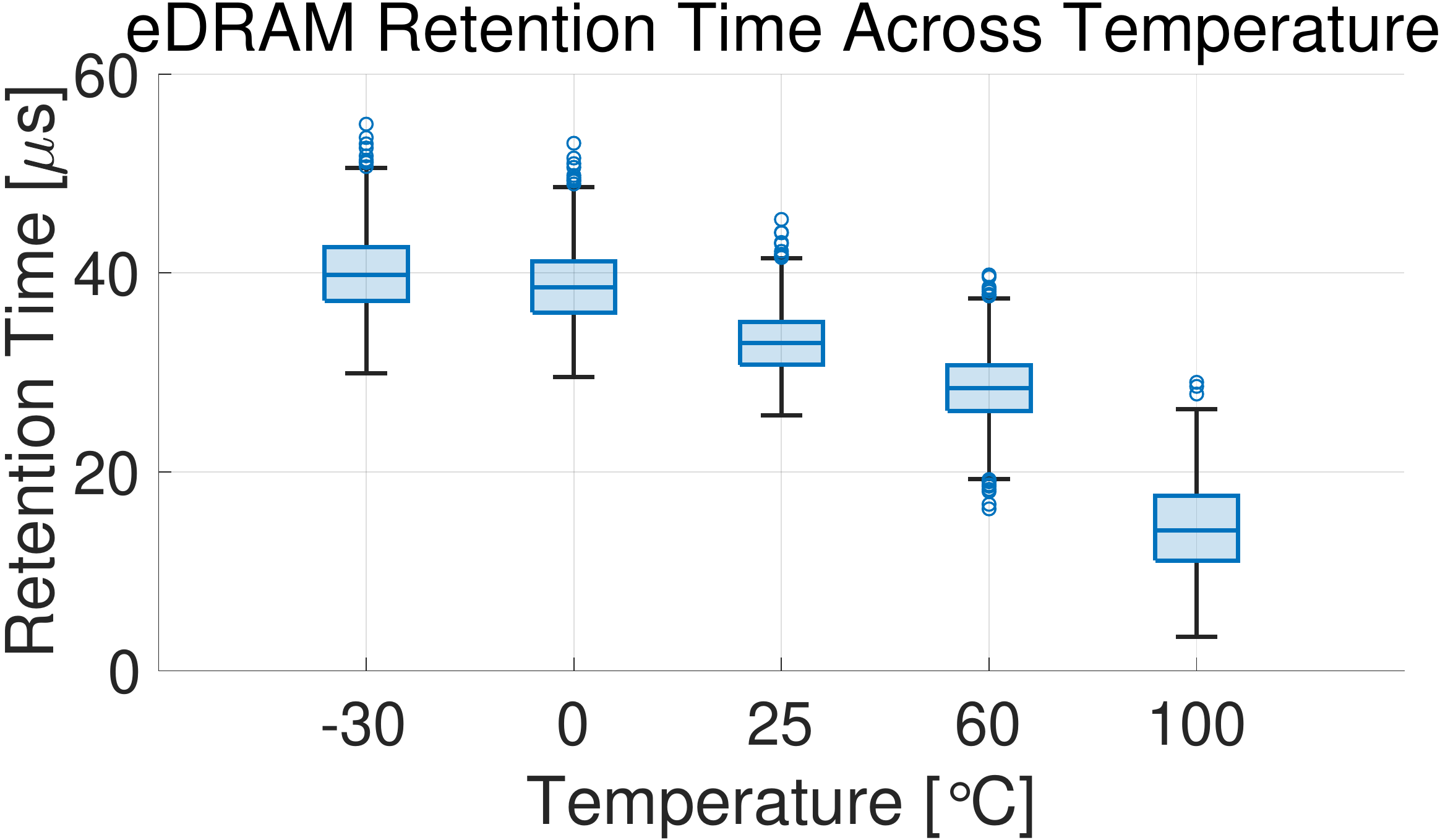}
    \caption{3T eDRAM retention time across temperature.}
    \label{fig:edram_rt}
    \squeezeup
\end{minipage}
\begin{minipage}{0.22\textwidth}
  \vspace{0pt}
    \centering
    \small
    \begin{adjustbox}{width=1\columnwidth,center}
    \begin{tabular}{c|c|c|c} \hline
    Datasets &  DuDNN  & FI & CA\\ \hline
    MRPC   &  0.822  &  0.83 &  0.787  \\
    WNLI   &  0.606  &  0.608 & 0.534  \\
    RTE    &  0.678  &  0.669 & 0.612  \\
    STS-B  &  0.879  &  0.889  & 0.768 \\
    QNLI   &  0.903  &  0.909  & 0.845  \\
    SST-2  &  0.919  &  0.921  & 0.868 \\
    QQP    &  0.911  &  0.911  & 0.870 \\
    \hline
    \end{tabular}
    \end{adjustbox}
    \captionof{table}{Training accuracies over different GLUE tasks.}
    \label{tab:text-result}
    \squeezeup
\end{minipage}
\end{figure}

To model the behavior of eDRAM, we simulated a 58 bits $\times$ 1024 words 3T eDRAM array to extract retention time and energy consumption metrics. To find an accurate representation of retention time, we ran 2000 Monte Carlo on-die variation points at the typical process corner using VDD = 0.8V, and at temperatures ranging from -30 to 100 \degree C. We measure retention time to be immediately after a write operation up until the sense amplifier is still able to correctly read the stored value at 99.9\% yield. 
In these retention time simulations, we consider a worst-case memory access activity of 0.5 by toggling the write bit line (See Figure~\ref{fig:edram_map}) at the system operating frequency, corresponding to the high memory access activity of CAMEL. 
As Figure~\ref{fig:edram_rt} depicts, the worst-case retention time ranges from 3.35$\mu$s at 100 \degree C to 30$\mu$s at -30\degree C. In the subsequent hardware simulation, we assume the eDRAM has a retention time of 3.35$\mu$s. To highlight the superiority of eDRAM over SRAM, we conduct a comparison of the memory array area required by both technologies with a fixed memory capacity of 1Mbit. With a commercial 16nm process node, SRAM occupies an area of 103,428 $\mu m^{2}$, while eDRAM occupies 51,714 $\mu m^{2}$, which is roughly $2\times$ smaller.
For the area breakdown of the CAMEL system, the memory subsystem and systolic array dominate the area, accounting for 42.5\% and 33.0\%, respectively. The special function unit and accumulator follow, occupying 12.5\% and 12.0\% of the total area.





\subsection{Training Performance Evaluation}
\label{sec:hardware-perf}
We now evaluate hardware performance of the CAMEL system described in Section~\ref{sec:camel-hardware} by comparing it against another DNN training systems implemented without eDRAM. We configure the baseline system such that it has the same total on-chip area as our CAMEL system. Specifically, the baseline system contains a $4 \times 4$ systolic array where each systolic cell can perform the dot product between two groups of nine BFP numbers within one cycle. The memory subsystem includes six 48kB SRAMs for storing activations and gradients, and two 24kB SRAMs for storing the weights. Additionally, it incorporates a 4GB DRAM, similar to the one utilized by CAMEL, serving as the second-tier buffer. It contains an identical special function unit and accumulator as CAMEL. We refer to this baseline design as SRAM-based system, as it only uses SRAM as on-chip buffer.

We implement the DNN models that generate the accuracies on Tiny-ImageNet shown in Table~\ref{tal:acc-eval} using the CAMEL and SRAM-based systems. We use Time-to-Accuracy (TTA)~\cite{coleman2017dawnbench} as the evaluation metric to compare the different approaches. TTA measures the amount of total latency required to train the DNN model until reaching a target validation accuracy. Moreover, we also measure the total amount of energy during this training process, and name the resulting metric~\textit{Energy-to-Accuracy} (ETA). In the case of Tiny-ImageNet, we aim for a target accuracy of 60\%, which is considered high for this particular dataset.

To clearly investigate the individual impact of DuDNN and eDRAM towards the overall system performance improvement, we compare the TTA and ETA of different DNN algorithms (DuDNN, FI, CA, BO) over CAMEL and the SRAM-based baseline system. This translates to eight evaluation candidates: 1. DuDNN implemented on CAMEL (DuDNN + CAMEL) 2. DuDNN implemented on the SRAM-based system (DuDNN + SRAM-based) 3. FI + CAMEL 4. FI + SRAM-based 5. CA + CAMEL 6. CA + SRAM-based. 7. BO + CAMEL 8. BO + SRAM-based. We set the mini-batch size to 64 for all of them for training.



\begin{figure}
    \centering
    \includegraphics[width=0.48\textwidth]{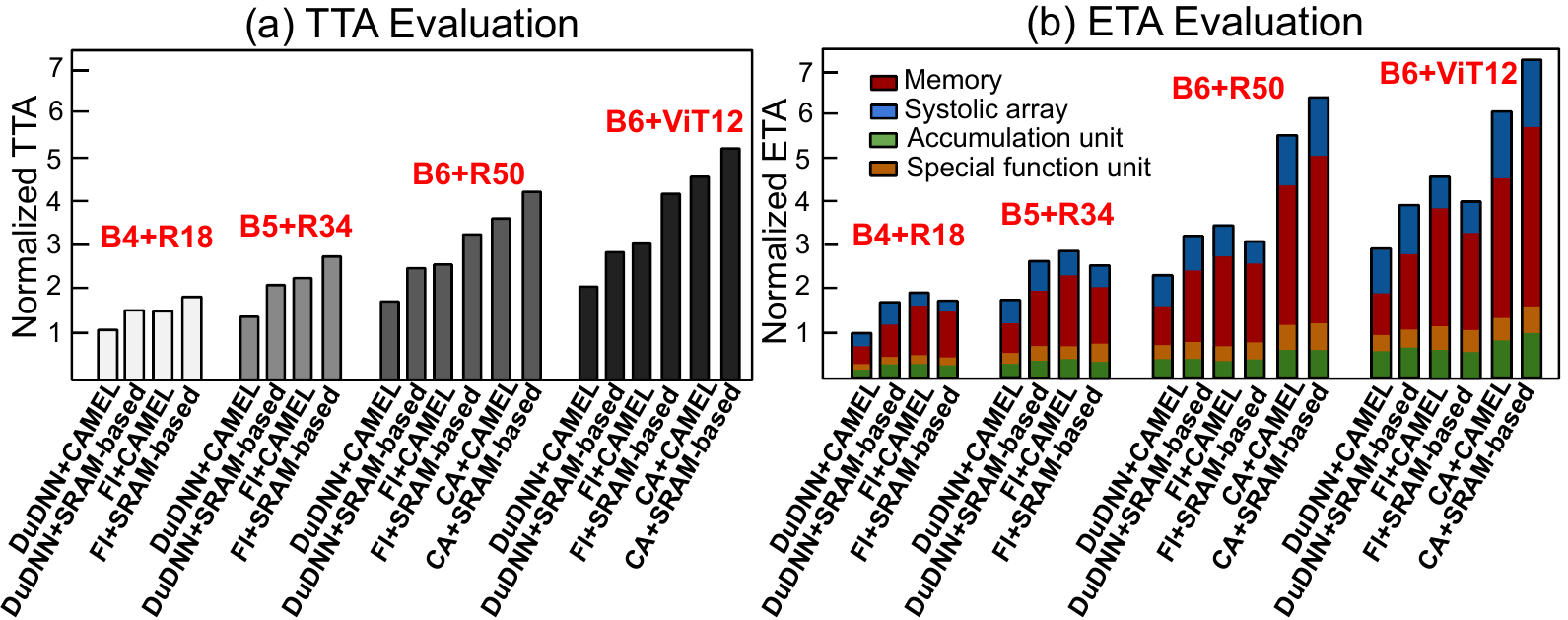}
    \caption{TTA and ETA evaluation over different DNN architectures on Tiny-ImageNet. "B" and "R" denote Branch and ResNet, respectively.}
    \label{fig:ttaeta}
    \squeezeup
\end{figure}

\subsubsection{Training performance on small DNN models}
Figure~\ref{fig:ttaeta}(a) illustrates the TTA of the baseline systems using various DNN architectures on Tiny-ImageNet. The training time is normalized based on the performance of DuDNN+CAMEL, which achieves the smallest TTA.  For certain settings, such as CA + SRAM-based, CA + CAMEL, BO+SRAM-based and BO+CAMEL, fail to reach the target accuracy of 70\% and are consequently excluded from the figure.  
For small DNN models, including Branch-4 + ResNet-18, Branch-5 + ResNet-34, Branch-6 + ResNet-50, and Branch-6 + ViT-12, all the intermediate activations and gradients in DuDNN can be accommodated entirely within the on-chip eDRAM. Moreover, these models' data lifetime throughout their training remains strictly below the worst eDRAM retention time of 3.35$\mu$s. As demonstrated in Table~\ref{tab:datalife-time}, which showcases the maximum data lifetime during the training process of each DNN model on Tiny-ImageNet, this leads to a refresh-free eDRAM during the entire training process of DuDNN + CAMEL. Additionally, we also show the average eDRAM utilization rate across a single iteration of training process. We observe that the eDRAM achieve an $89\%-93\%$ utilization rate across different DNN architectures.
\begin{table}
\centering
\caption{Data lifetime during the DuDNN training. "B", "R" and "V" denote Branch, ResNet and VGG, respectively.}
\begin{adjustbox}{width=\columnwidth}
\begin{tabular}{l|c|c|c|c} \hline
DNNs       &  B4+R18  & B5+R34 & B6+R50 & B6+ViT12 \\ \hline
lifetime   &    2.66$\mu$s  & 2.87$\mu$s   &  3.03$\mu$s  &   3.10$\mu$S \\ \hline
\#Refresh  &   0    &   0  &   0  &    0 \\\hline
Util. rate (\%)  &   88.9\%    &   90.0\%  &   91.7\%  &    92.2\% \\\hline\hline
DNNs    & B6+V16 & B8+ViT16 & B12+R50 & B12+ViT16 \\\hline
lifetime &   3.86$\mu$s &  4.02$\mu$s    & 3.40$\mu$s  & 4.13$\mu$s   \\\hline
\#Refresh     & 1 & 1 & 1 & 1   \\\hline
Util. rate (\%) &  92.4\%    &   92.1\%  &   91.5\%  &    92.8\% \\\hline
\end{tabular}
\end{adjustbox}
\label{tab:datalife-time}
\end{table}

We make the following observations from the figure. First, DuDNN + CAMEL achieves optimal performance over all other solutions. Compared with eDRAM, the SRAM-based system utilizes SRAM to store all activations and gradients on chip. In contrast to eDRAM, SRAM exhibits lower density, resulting in a larger area consumption. Consequently, this reduces the size of the systolic array, leading to lower throughput compared to CAMEL. Despite the fact that DuDNN training involves an additional process of recomputing the input, it still achieves an average TTA that is $2.2\times$ smaller on average.
Furthermore, when compared to DuDNN, CA and BO experience inferior convergence behavior, which requires many more iterations to achieve the target accuracy, resulting in significantly higher TTA. Both FI+SRAM-based and FI+CAMEL methods demand the storage of intermediate data during the forward pass. Consequently, this imposes substantial memory footprint and high communication overhead between the on-chip buffer and DRAM, leading to longer training latency.

Figure~\ref{fig:ttaeta}(b) illustrates the ETA evaluation of various approaches. We see that DuDNN+CAMEL achieves more than $2.8\times$ lower ETA on average compared with the other approaches. This is partially due to the fact that DuDNN+CAMEL obtains the smallest TTA among all of the systems, which further leads to the smallest energy consumption. Furthermore, the integration of eDRAMs in CAMEL results in significantly lower memory access and leakage power compared to SRAM-based systems. Despite the power increase caused by the input recomputation in DuDNN, the benefits of using eDRAM outweigh this growth, leading to an overall reduction in power consumption. When comparing DuDNN+CAMEL to DuDNN+SRAM-based systems, the former achieves an average reduction of $2.34\times$ in ETA across different DNN architectures. This highlights the advantage of eDRAM over SRAM for DuDNN training. The reversibility of DuDNN results in minimal on-chip storage with short data lifetime, significantly reducing eDRAM refresh costs and leading to lower memory power consumption. However, employing eDRAM for training irreversible DNN architectures (e.g., FI) proves inefficient. This is because FI requires to retain all intermediate data in eDRAM for a long period, leading to a large number of refresh operations and increased energy consumption.
\begin{table}
 \centering
 \caption{The TTA/ETA of each DNN running on CAMEL and SRAM-based systems. }
\begin{adjustbox}{width=\columnwidth,center}
    \begin{tabular}{lcccccccc}\toprule 
    & \multirow{2}{*}{\begin{tabular}[c]{@{}c@{}}Methods \end{tabular} }
    & \multicolumn{2}{c}{B8+ViT16} & \multicolumn{2}{c}{B12+R50} & \multicolumn{2}{c}{B12+ViT16}  \\
    \cmidrule(lr){3-4} \cmidrule(lr){5-6} \cmidrule(lr){7-8}
    &  & CAMEL  & SRAM-based  & CAMEL  & SRAM-based  & CAMEL  & SRAM-based  \\ \toprule 
    & DuDNN   & $1\times$/$1\times$ & $1.25\times$/$1.39\times$  & $1.81\times$/$1.86\times$ & $2.00\times$/$2.28\times$   & $2.51\times$/$2.73\times$ & $3.33\times$/$3.91\times$ \\
    & FI   & $1.27\times$/$1.47\times$ & $1.30\times$/$1.32\times$  & $1.95\times$/$2.49\times$  & $2.20\times$/$2.15\times$  & $3.48\times$/$4.20\times$ & $3.62\times$/$3.72\times$ \\
    & CA   & inf/inf & inf/inf & $3.45\times$/$3.83\times$ & $4.33\times$/$4.76\times$ & $5.20\times$/$5.75\times$ & $6.21\times$/$6.64\times$  \\
    & BO   & Inf/Inf & Inf/Inf & $7.82\times$/$8.32\times$ & $8.43\times$/$9.04\times$  & $8.47\times$/$8.86\times$  & $9.03\times$/$9.34\times$  \\
\midrule
    \end{tabular}
\end{adjustbox}
\label{tal:eta-tta-large}
\end{table}
\subsubsection{Training performance on large DNN models}
For larger DNN models, such as Branch-8 + ViT-16, Branch-12 + ResNet-50, and Branch-12 + ViT-16, the activations and gradients in the DuDNN training require transfer and storage in the off-chip DRAM. Due to the substantial model size, the data lifetime and eDRAM refresh power increase during the DuDNN training. The maximum data lifetime for each DNN model during the training process is presented in Table~\ref{tab:datalife-time}. From the data, we can observe that the eDRAM requires only one refresh to preserve the integrity of the stored data. Table~\ref{tal:eta-tta-large} displays the TTA and ETA of training each DNN model over the CAMEL and SRAM-based systems. The entries are in the format of TTA/ETA. For methods that cannot reach the target accuracy (70\%), both TTA and ETA are denoted as infinity (Inf). We notice that a similar trend as Figure~\ref{fig:ttaeta}. This demonstrates that DuDNN+CAMEL can achieve superior training performance across various DNN architectures with different sizes.  
\begin{table}
\centering
\caption{Performance of DuDNN+CAMEL with and without refresh.}
\begin{adjustbox}{width=\columnwidth,center}
\begin{tabular}{c|c|c|c|c} \hline
DNNs       &  B4+R18  & B5+R34 & B6+R50 & B6+ViT12 \\ \hline
\#Refresh=0 &   $1\times$   &   $1\times$  &   $1\times$  &    $1\times$ \\\hline
\#Refresh=1 &   $1.10\times$   &   $1.13\times$  &   $1.15\times$  &    $1.18\times$  \\\hline
\end{tabular}
\end{adjustbox}
\label{tab:robustness}
\end{table}

\subsection{Impact of eDRAM refresh on ETA}
We present the performance variation of DuDNN + CAMEL with and without eDRAM refresh during the training process. Specifically, we evaluate the ETA of DuDNN + CAMEL for three configurations: Branch-4 + ResNet-18, Branch-6 + ResNet-50, and Branch-6 + ViT-12. We achieve this by increasing the number of eDRAM refreshes, even though the simulation results indicate that eDRAM refresh is unnecessary. All other settings are maintained constant. As shown in Table~\ref{tab:robustness}, the ETA for these three DNNs moderately increases with a higher number of refreshes. However, it still remains superior to or on par with the performance of the SRAM-based design. 

\subsection{Impact of eDRAM refresh on accuracy}
\begin{table}
\centering
\caption{Impact of eDRAM refresh on FI training accuracy.}
\begin{adjustbox}{width=\columnwidth,center}
\begin{tabular}{|l|c|c|c|c|}  \hline
Datasets &  B4+R18  & B6+R50 & B8+ViT16 & B12+ViT16 \\ \hline
CIFAR-10    &   23.23\%  &    16.67\%  &    12.03\%   &    10.05\%   \\
Tiny-ImageNet &  1.5\%   &    0.88\% &   0.5\%   &    0.5\%  \\\hline
\end{tabular}
\end{adjustbox}
\label{tab:refresh-accu}
\end{table}
To highlight the significance of the eDRAM refresh operation DNN training accuracy, we measured the training accuracy of FI+CAMEL by intentionally excluding the eDRAM refresh operation during the training phase. This elimination results in the data stored for gradient computation during the backward process becoming corrupted due to the extended data lifetime during FI training, leading to a significant accuracy degradation. This is supported by the results presented in Table~\ref{tab:refresh-accu}, which clearly indicate a substantial drop in accuracy when compared to the FI accuracy reported in Table~\ref{tal:acc-eval}.

\subsection{Accuracy Performance with Energy}
\label{sec:accuracy-energy}
\begin{figure}
\centering
\begin{minipage}[t]{0.25\textwidth}
  \vspace{0pt}
    \includegraphics[width=0.87\textwidth]{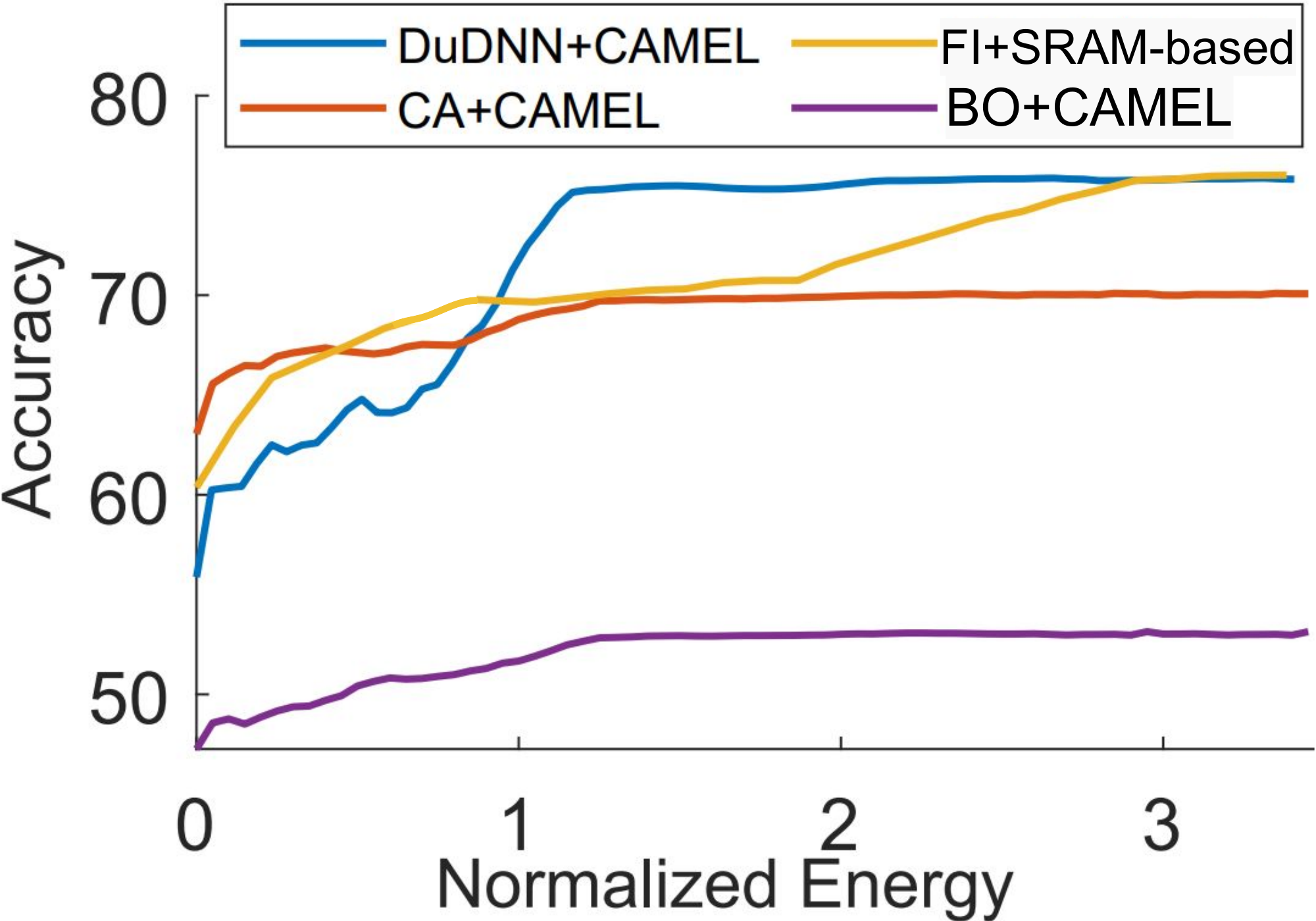}
    \caption{Accuracy versus energy.}
    \label{fig:acc-energy}
    \squeezeup
\end{minipage}\hfill%
\begin{minipage}[t]{0.23\textwidth}
  \vspace{0pt}
  \small	
\begin{tabular}{| c | c |} \hline
Array size & Data Lifetime\\\hline
$6\times 6$ & $1\times$ \\ \hline
$10\times 10$ & $0.90\times$ \\\hline
$12\times 12$ & $0.37\times$ \\\hline
\end{tabular}
\captionof{table}{Impact of systolic array size on data lifetime during the training of Branch-6+ResNet-50. The data lifetime are normalized.} 
\label{tab:array-eta}
\squeezeup
\end{minipage}\hfill%
\end{figure}

Figure~\ref{fig:acc-energy} shows the changes in model accuracies versus energy consumption for the different approaches with Branch-6+ResNet-50 architecture on Tiny-ImageNet dataset. We see that DuDNN+CAMEL converges with the least amount of energy. By comparison, FI+SRAM-based ends up with slightly higher accuracy than DuDNN+CAMEL at a price of much larger total energy consumption. Finally, CA+CAMEL and BO+CAMEL converge at much lower accuracy with much higher energy consumption.
\subsection{Impact of Systolic Array Size}
\label{sec:systolic-array-impact}
Finally, we investigate the impact of the systolic array size on data lifetime. Specifically, we increase the systolic array size as well as the number of eDRAM and SRAM banks proportionally to support systolic array operation. Table~\ref{tab:array-eta} presents the average data lifetime during the training process of Branch-6+ResNet-50 on Tiny-ImageNet. We observe that the data lifetime decreases sub-linearly as the systolic array size grows. This sublinearity is because a larger systolic array may also incur a lower utilization rate for some layers of DuDNN, resulting in the same retention time as the small systolic array.

\section{Conclusion and Future Work}
\label{sec:conclusion}
In this work, we propose a novel DNN training system that uses eDRAM as the major storage medium of transient data. To mitigate the expensive data refresh in eDRAM, we further propose a novel Duplex DNN architecture that enables a significantly reduced data lifetime during the DNN training process and further eliminates the needs for eDRAM refreshing. 
Finally, we propose an optimal computation pattern for Duplex DNN training with minimized memory consumption and data lifetime. 
To our best knowledge, this is the first work that proposes a model-scheduler-architecture stack for efficient on-chip training with refresh-free eDRAMs, which opens up interesting future avenues in a promising direction of research.
\newpage
\bibliographystyle{IEEEtranS}
\bibliography{refs}

\end{document}